\documentclass[a4paper,11pt]{article}
\pdfoutput=1 % if your are submitting a pdflatex (i.e. if you have
\usepackage{jcappub} % for details on the use of the package, please
\usepackage[T1]{fontenc} % if needed
\usepackage{multirow}
\usepackage{longtable}
\usepackage{tabularx}
\usepackage{booktabs}
\graphicspath{{plots/}}                      
\usepackage{ifthen}
\usepackage{hyperref}
\usepackage{url}
\usepackage{mathrsfs}
\usepackage{amsmath, amssymb}
\usepackage{subcaption} 
\usepackage{titlesec}
\usepackage{array}% 
\usepackage{float}
\usepackage[normalem]{ulem}

\def\nn{\nonumber}
\def\Ord{{\cal O}}
\def\pa{\partial}
\begin{document}
\title{\boldmath Particle production during inflation:
	A Bayesian analysis with
	{CMB data from \textit{Planck} 2018}\\
	%\green{KF: Should 2018 be mentioned?} \magenta{PC: I think without 2018 is okay.}
}

\author[1]{Suvedha Suresh Naik,}
\author[1]{Kazuyuki Furuuchi,}
\author[2]{Pravabati Chingangbam}
\affiliation[1]{Manipal Centre for Natural Sciences,
Centre of Excellence, Manipal Academy of Higher Education,
Dr. T.M.A. Pai Planetarium Building, Manipal 576 104, Karnataka, India}
\affiliation[2]{Indian Institute of Astrophysics,
	Koramangala II Block, Bangalore 560 034, India}

% e-mail addresses: one for each author, in the same order as the authors
\emailAdd{suvedha.nail@learner.manipal.edu}
\emailAdd{kazuyuki.furuuchi@manipal.edu}
\emailAdd{prava@iiap.res.in}

%%%%%%%%%%%%%%%%%%%%%%%%%%%%%%%%%%%%%%%%%%%%
\abstract{
	A class of inflationary models that involve
	rapid bursts of particle productions predict
	observational signatures, such as bump-like 
	features in the primordial scalar power spectrum. 
	In this work, we analyze such models by
	comparing their predictions with the latest CMB data 
	from \textit{Planck} 2018.
	We consider two scenarios of particle production. 
	The first one is a simple scenario consisting of a single burst
	of particle production during observable inflation. 
	The second one consists of multiple bursts of particle production
	that lead to a series of bump-like features in the primordial power spectrum.
	We find that the second scenario of the multi-bump model 
	gives better fit to the CMB data %in the low-$\ell$ region
	compared to the concordance $\Lambda$CDM model. 
	We carried out model comparisons using Bayesian evidences.
	%Comparison of Bayesian evidence shows that 
	%these models are statistically indistinguishable 
	%from the concordance model with the current CMB data. %\magenta{$<--$ Shall we remove this sentence. I wonder if it weakens the abstract}
	%We then obtain a bound on one of the theoretical model parameters 	using the observational constraints on the amplitude of primordial features. \magenta{$<--$ How about we change this sentence to this $->$} 
	From the observational constraints on the amplitude of primordial features of the multi-bump model, we find that the dimensionless coupling parameter $g$ responsible for particle production is bound to be $g< 0.05$.
	}

\maketitle
%%%%%%%%%%%%%%%%%%%%%%%%%%%%%%%%%%%%%%%%%%%
\section{Introduction}
\label{sec:intro}
%%%%%%%%%%%%%%%%%%%%%%%%%%%%%%%%%%%%%%%%%%%%
%\noindent\magenta{POINT 1 : General introduction to inflation}
Cosmic inflation~\cite{Guth:1981,LINDE:1982,Albrecht:1982,STAROBINSKY198099,stato:1981,kazanas:1980} is the hypothetical accelerated expansion in the early universe,
which was proposed to explain the unnatural initial conditions 
of the Big Bang model. The relevant energy scale of the inflationary era is so high that any observations are beyond the reach of direct exploration. 
However, modeling cosmic inflation and predicting its signatures in 
cosmological observations provides an alternative indirect strategy to understand the physics at such high energy scales.
%together with what happened in the early universe. 
The observations of density fluctuations provide valuable information about the history of the universe. According to the inflationary scenario, the seeds of the density fluctuations were supposed to have been created during inflation.
Hence, the fluctuations in the distribution of matter and radiation that we observe today are expected to contain important clues about how inflation happened. Due to the high energy scale involved, the cosmic inflation must be described by a  particle physics model beyond the Standard Model.

%\noindent\magenta{POINT 2: Inflation models and how data constrain model parameters, possibility of existence of features}
A plethora of inflationary models have been proposed over the years and tested using cosmological observations such as Cosmic Microwave Background (CMB) radiation and matter distributions.
The tests provide constraints on the parameters of the underlying theoretical models. 
Cosmological observations have suggested that our universe can be modeled accurately by  the concordance $\Lambda$CDM model. 
This model successfully explains a wide range of cosmological phenomena with just six parameters. 
The concordance model assumes an almost scale-invariant primordial power spectrum of scalar perturbations characterized by only two parameters: 
the amplitude of the  perturbations and the spectral tilt of the power spectrum. 
Any hint of disagreement between the concordance model and
observations may indicate the presence of primordial features or the potential signatures of new physics in the early universe.
Therefore, more precise observations are required to distinguish between the concordance and alternate models. 
\textit{Planck} \cite{Akrami:2018vks} gives the best available full-sky CMB  data at present. 
%
%~\cite{Chen:2008wn,Silverstein:2008sg,Flauger:2009ab,Adams:2001vc,Bean:2008na,Achucarro:2010da,Miranda:2012rm,Bartolo:2013exa,Hazra:2014goa,Green:2009ds,Green:2014xqa,Amin:2015ftc,Garcia:2019icv,Barnaby:2009dd,Chantavat:2010vt,Chen:2011zf,Chen:2014joa,Chen:2015lza,Chen:2018cgg,Starobinsky:1992ts,Ivanov:1994pa,Covi:2006ci,Ashoorioon:2006wc,Joy:2007na,Joy:2008qd,Hazra:2010ve,Benetti:2013cja,GallegoCadavid:2014jac,Chluba:2015bqa,Bousso:2014jca,Allahverdi:2006iq,Jain:2008dw,McAllister:2008hb,Pahud:2008ae,Flauger:2010ja,Chen:2010bka,Aich:2011qv,Peiris:2013opa,Meerburg:2013dla,Easther:2013kla,Motohashi:2015hpa,Miranda:2015cea,Cremonini:2010ua,Braglia:2020fms,Braglia:2020eai,Chen:2012ja,Chen:2014cwa,Braglia:2020taf,Braglia:2021ckn,Hazra:2016fkm,Hazra:2014jka,Hazra:2017joc,LHuillier:2017lgm,Debono:2020emh}.
\textit{Planck}'s residuals of the CMB power spectrum with respect to 
the concordance model show
hints of anomalies at different angular scales.
A detailed parametric search for features 
in the primordial scalar power spectrum  was performed using \textit{Planck} data in \cite{Akrami:2018odb}
to interpret the anomalies. 
After exploring several classes of inflationary models that predict primordial features, 
Ref.~\cite{Akrami:2018odb} concluded that no statistically significant evidence for the features is present in the CMB data. 
%Nevertheless, the data do not entirely rule out the possibility  of the presence of features. 

%\noindent\magenta{POINT 4 : Other types of features in the literature}
Searches for features in the primordial power spectrum have attracted considerable interests during the past two decades. 
Here, we provide a non-exhaustive list of various classes of models that produce primordial features.
A class of models predicts oscillatory
components for the power spectrum over the entire observable
range of co-moving wave-numbers. 
A well-known example is the axion monodromy model
\cite{Silverstein:2008sg,Flauger:2009ab}.
Models with a sharp feature in the inflaton potential leading
to localized oscillations in the power spectrum
have been studied in
\cite{Adams:2001vc,Chen:2006xjb,Achucarro:2010da,Miranda:2012rm,Bartolo:2013exa,Hazra:2014goa,Jain:2008dw};
a kink in the potential gives rise to power suppression at the largest scales \cite{Starobinsky:1992ts}.
Multi-field inflation models can also generate 
sharp or resonant features in the primordial power spectrum
\cite{Cremonini:2010ua,Braglia:2020fms,Braglia:2020eai,Chen:2014cwa,Braglia:2021ckn,Braglia:2021sun,Braglia:2021rej}.
%lpha attractor multi-field models generate enhanced powers on small-scales \cite{Iacconi:2021ltm}.
Although previous studies show that 
the Bayesian analyses favor the concordance model 
over other inflationary models producing primordial features, 
the search for primordial features is well-motivated 
from the perspectives of the theory as well as 
current and future observations.

%\noindent\magenta{POINT 3 : features from theoretical models, the model of interest here, and what is the goal of this paper}
%Some particle physics models of inflation naturally predict deviations from the featureless power spectrum of fluctuations assumed by the concordance cosmology model. 
In this paper, we focus on a class of inflation models that involves bursts of particle production during inflation
%, which may leave observable signatures such as features in the primordial density perturbations 
\cite{Chung:1999ve,Barnaby:2009mc,Barnaby:2009dd,Barnaby:2010ke,Chluba:2015bqa,Pearce:2017bdc,Ballardini:2022wzu}.
It was recently shown that such bursts of particle productions naturally occur in inflation models based on higher-dimensional gauge theories \cite{Furuuchi:2015foh,Furuuchi:2020klq,Furuuchi:2020ery},
leading to renewed interest in searching for the corresponding features in the observed data of primordial density perturbations. This class of models predicts that  bursts of particle production associated with the inflaton motion give  rise to bump-like features in the power spectrum. 
We investigate the presence of such features 
%motivated by the particle physics models of inflation~\cite{Furuuchi:2015foh,Furuuchi:2020klq,Furuuchi:2020ery}
by confronting them with the \textit{Planck} data. 
For comparison with observed data, the resulting primordial power spectrum can be modeled by suitable analytical expressions for the shape and height of the bump. 
In \cite{Barnaby:2009dd}, the functional form of the power spectrum for bump-like features was obtained by fitting the numerical results.
%Ref.~\cite{Barnaby:2009dd} gave a power spectrum template for the bump-like features using numerical fits. 
Here, we use the analytical power spectrum
(one-loop approximation) given in~\cite{Pearce:2017bdc}. 
%Their results include contributions to the power spectrum,
Their result shows that, 
at co-moving wave-numbers $k$ greater than those of peak,
the contributions to the power spectrum is characterized by oscillations, 
modulated by an amplitude that decreases as $k^{-3}$,
while \cite{Barnaby:2009dd}
showed an exponentially decreasing function.
It should be noted that the power spectrum template 
used in our analysis
%predicted by these models 
%(analytical form of the contributions given in \cite{Pearce:2017bdc})
is different from the templates
investigated in \cite{Akrami:2018odb}. 
%To our knowledge, 
%this work is the first observational analysis 
%of primordial features with the form given in \cite{Pearce:2017bdc}.

This article is organized as follows: In section~\ref{sec:model}, we briefly describe the inflation model involving particle productions during inflation and its observational predictions on the power spectrum. 
Section~\ref{sec:method} discusses our methodology of constraining
the model parameters with CMB data. 
In section~\ref{sec:results}, we 
present the results of Bayesian analyses for both the single bump and multi-bump models. 
Finally, we summarize our work and discuss perspectives for future work in section~\ref{sec:summary}. 
%%%%%%%%%%%%%%%%%%%%%%%%%%%%%%%%%%%%%%%%%%%%%%%%%
%
%%%%%%%%%%%%%%%%%%%%%%%%%%%%%%%%%%%%%%%%%%%%%%%%%%%%%%%%%%
\section{The Models}
\label{sec:model}
%%%%%%%%%%%%%%%%%%%%%%%%%%%%%%%%%%%%%%%%%%%%
In this section, we briefly describe the inflation models we will analyze.
We will study a class of models
in which a real massless\footnote{%
In the current context, fields whose
mass is much lighter than the Hubble scale
at the time
can be treated as massless fields.}
scalar field $\chi$
couples with the inflaton field $\phi$ 
via the interaction term
\begin{equation}
g^2 (\phi-\phi_0)^2 \chi^2\,,
\label{eq:masschi}
\end{equation} 
where $g^2$ is a dimensionless coupling constant. 
When the inflaton field value crosses 
$\phi = \phi_0$, 
the $\chi$ particles become instantaneously massless,
and this results in bursts of $\chi$ particle production. 
%\magenta{Does this happen by parametric resonance? 
%Perhaps we should some more explanation of how the particle production happens.} 
%\green{KF: Pearce et al's summary is adequate, we may follow.}

% 
Assuming that this mechanism occurs during the observable 
range of e-folds of inflation, 
the primordial power spectrum may 
accommodate a series of bump-like features.
%\cite{Barnaby:2009dd,Barnaby:2009mc,Chung:1999ve,Pearce:2017bdc}.
%
%
%In some of the recently proposed
Furthermore, it has recently been pointed out that
inflation models based on gauge theory
in higher dimensions naturally give rise to 
the coupling of the form eq.~\eqref{eq:masschi}
\cite{Furuuchi:2015foh,Furuuchi:2020klq,Furuuchi:2020ery}.
Here, we write down the low energy action appropriate for
describing inflation 
arising from these models:
\begin{align}
S
=
\int d^4 x
 \Biggl[
% &- \frac{1}{4} F_{\mu\nu} (x) F^{\mu\nu} (x)
% +
% \frac{m^2}{2}
% \cA_\mu (x) \cA^\mu (x)
% \Biggr.
% \nn\\
% &
% +
&\frac{1}{2} \pa_\mu \phi (x) \pa^\mu \phi (x)
- V(\phi)
\nn\\
\Biggl.
&+
\sum_{n=-\infty}^{\infty}
\left\{ 
D_\mu {\chi}_n^{\dagger} (x) D^\mu \chi_n (x)
+
\chi^\dagger_{n} (x)
g^2
\left(
\phi(x) - 2\pi f n
\right)^2
\chi_n (x)
\right\}
\Biggr]\,.
\label{eq:S4}
\end{align}
Here, $\phi$ is the inflaton field
that originates from the extra-dimensional
component of the gauge field,
$f$ is the symmetry breaking scale,
and $n$ takes integer values.
$V(\phi)$ is the potential for the inflaton field.
$\chi_n$'s are complex scalar fields
charged under the gauge group
(here, for simplicity, we consider $U(1)$ gauge group.).
$D_\mu$ stands for the gauge covariant derivative.
In the models based on higher-dimensional gauge theory,
$g$ is the 4D gauge coupling constant related to the 
gauge coupling in higher dimensions.
Like the interaction \eqref{eq:masschi},
when the inflaton crosses
the value $2\pi f n$,
the field $\chi_n$ becomes massless,
leading to the burst of 
$\chi_n$ particle production.\footnote{%
In models in which the 
extra dimensions are deconstructed ones
\cite{Furuuchi:2020klq,Furuuchi:2020ery},
the interaction terms
between the inflaton and the complex scalar fields
are slightly different,
but the effects on the particle productions
are similar.}
In the models based on higher-dimensional gauge theory,
this interaction originates from the dimensional reduction
of the gauge covariant kinetic term (minimal coupling)
in higher dimensions.
A distinctive feature of the models
based on higher dimensional gauge theory
is that 
a scalar field becomes massless
periodically with respect to the value of $\phi$.
%*** after talking about the power spectrum ***
However, the bump-like features in the power spectrum
do not equally spread in the co-moving scale space.
Let us parametrize the separation between the $i$-th
and $(i+1)$-th bumps as
\begin{equation}
k_{i+1} = e^{\Delta(\phi)} k_i \,,
\end{equation}
where $\Delta(\phi)$ is estimated as
\begin{equation}
\Delta(\phi)
\simeq \frac{dN}{d\phi}(\phi)  2\pi f 
\simeq \frac{1}{M_P^2} \frac{V(\phi)}{V'(\phi)} 2 \pi f \,.
\label{eq:Delta}
\end{equation}
Here, $N(\phi)$ is the number of e-folds:
\begin{equation}
N(\phi) 
\simeq 
\frac{1}{M_P^2}
\int_{\phi_{end}}^{\phi} d\varphi
\,
\frac{V(\varphi)}{V^\prime(\varphi)} \,,
\label{eq:Ne}
\end{equation}
where $\phi_{end}$
is the value of the inflaton at the end of inflation.\footnote{%
More concretely, we define $\phi_{end}$ as the value of the inflaton field
when the slow-roll parameter $\epsilon$ reaches one.}

As seen from eq.~\eqref{eq:Delta},
the spacing between
the $i$-th and $(i+1)$-th bumps
is a function of the inflaton field value
when those co-moving wavenumbers $k_i$ and $k_{i+1}$ 
exited the horizon.
However, this $\phi$-dependence is mild
in the cases of our interests:
Let $\delta \phi$ be the change
in the inflaton field values
that correspond to the range of scales that
exit the horizon, and that can be observed by CMB.
%corresponding to the largest
%scales to the smallest scales observable with CMB 
%exited the horizon.
The relative change of $\Delta$ 
due to the change in $\delta \phi$ is estimated as
\begin{equation}
\frac{\delta \Delta}{\Delta}
\simeq
\frac{\frac{d^2 N}{d\phi^2} \delta \phi}{\frac{d N}{d\phi}}
\simeq
\sqrt{2\epsilon}
\left(
1-\frac{\eta}{2\epsilon}
\right)
\frac{\delta \phi}{M_P}\,.
\label{eq:deltaDelta}
\end{equation}
Here, $\epsilon$ and $\eta$ are the slow-roll parameters:
\begin{equation}
\epsilon(\phi) 
=
\frac{M_P^2}{2}
\left(\frac{V'}{V}\right)^2
	\,,
\qquad
\eta(\phi) 
= 
M_P^2 \frac{V''}{V}  \,.
   \label{eq:slowrollps}
\end{equation}
In typical large field inflation models,
%eq.~\eqref{eq:deltaDelta}
$\frac{\delta \Delta}{\Delta}$
is of $\Ord(10^{-1})$.
For example, in the case of monomial potential
$V(\phi) = \lambda \phi^p/p!$,
\begin{equation}
\frac{\delta \Delta}{\Delta}
\simeq
\frac{\delta \phi}{\phi}
\lesssim \Ord(10^{-1})\,.
\label{eq:monoDelta}
\end{equation}
We will later analyze the power spectrum with constant $\Delta$.
It turns out that in the parameter region of interest, 
$\Delta \ll 1$, and
the $\phi$-dependence of $\Delta$ does not alter
the qualitative feature of the power spectrum significantly.
In this parameter region, 
the bump-like features 
in the power spectrum
largely overlap,
and 
the contribution of the particle productions
to the power spectrum 
is not clearly distinguishable from
a constant contribution.

%%%%%%%%%%%%%%%%%%%%%%%%%%%%%%%%%%%%%%%%%
% \subsection{Feedback to model building}
% *** will think where to put later ***
%%%%%%%%%%%%%%%%%%%%%%%%%%%%%%%%%%%%%%%%

In the original models based on higher-dimensional gauge theory
\cite{Furuuchi:2015foh,Furuuchi:2020klq,Furuuchi:2020ery},
there was no particular co-moving scale
from which the bump-like feature starts:
The particle productions
and resulting bump-like features
were present
throughout the inflation period.
Later on, we will be motivated to
extend these models
%so that they
to produce
bump-like features starting at a certain co-moving scale $k_1$.
This can be achieved by 
adding another inflaton-dependent mass
to the $\chi_n$ fields,
so that before the co-moving scale
$k_1$ exits the horizon
$\chi_n$'s are massive,
while after 
the co-moving scale
$k_1$ exits the horizon
they become massless.
For example, we can add another inflaton-dependent mass term
of the form\footnote{Without a loss of generality, we assume $\phi$ decreases with time during inflation.}
\begin{equation}
m^2(\phi) \chi_n^\dagger \chi_n \,,
\qquad
m^2(\phi) = m^2 \left( 1 - \tanh (\phi_1 - \phi)  \right)\,.
\label{eq:massfn}
\end{equation}
The functional form of $m^2(\phi)$ in eq.~\eqref{eq:massfn}
is 
just for an illustrative purpose.
Any function which 
makes $\chi_n$
massive before $k_1$
and massless after $k_1$
will serve our purpose.

%
% and the resulting scalar power spectrum 
% would be different from
% the usual power-law  form assumed in the $\Lambda$CDM model.
%
%%%%%%%%%%%%%%%%%%%%%%%%%%%%%%%%%%%%%%%%%%%%%%%%%%%%%%%%%%%%%%%%%%
\subsection*{Power spectrum of primordial density perturbations}
%%%%%%%%%%%%%%%%%%%%%%%%%%%%%%%%%%%%%%%%%%%%%%%%%%%%%%%%%%%%%%%%%%
\textbf{The fiducial model:} 
The concordance $\Lambda$CDM model assumes a scalar power spectrum parametrized by
the amplitude of the scalar perturbations $A_s$
and scalar spectral index $n_s$:
%The scalar power spectrum for the power law is given by
\begin{equation}\label{eq:PS_PowerLaw}
P_s(k)
=
A_s \left(\frac{k}{k_\ast}\right)^{n_s-1}\,,
\end{equation}
where $k$ is the co-moving wave-number and the pivot scale $k_\ast$
is %\green
{chosen to be} $0.05~\text{Mpc}^{-1}$ in~\cite{Akrami:2018vks}.
Current data from 
\textit{Planck}\footnote{Combining the temperature, 
	polarization, and lensing data from the 2018 release.} 
determine the scalar spectral index in the $\Lambda$CDM model as
$n_s = 0.9649 \pm 0.0042$ 
%\magenta{at 68\% or 95\% confidence level (CL)?}
at $68\%$ confidence level. 
%making the power spectrum nearly scale-invariant. 
We will refer to this power spectrum template 
as the \textit{fiducial} model.\\[1mm]
\noindent \textbf{Models with primordial features due to particle production during inflation:} 
We parameterize the 
primordial power spectrum with features produced due to  
particle production during inflation.
%A parametrization of the 
%primordial power spectrum with features produced due to  
%particle production during inflation
%was given in \cite{Barnaby:2009dd}:
We include the dominant and subdominant
contributions to the power spectrum, 
calculated analytically with one-loop approximations \cite{Pearce:2017bdc}:
\begin{equation}
P_s(k)
=
A_s \left(\frac{k}{k_\ast}\right)^{n_s-1}
+
A_{\rm I} \sum_i \left(\frac{f_1(x_i)}{f_1^{\rm max}}\right)
+
A_{\rm II} \sum_i \left(\frac{f_2(x_i)}{f_2^{\rm max}}\right)\,,
\label{eq:PS_bump}
\end{equation}
where 
$A_{\rm I}$ and $A_{\rm II}$
are amplitudes of dominant and
subdominant contributions, respectively.
The amplitudes depend on the model parameter $g$ as\footnote{The dependence of $A_{\rm I}$ and $A_{\rm II}$ on $g$ is 
%considered for one real degree of freedom.
the case of a real scalar field.
A factor of two should be multiplied appropriately for the case of a complex scalar field.}
\begin{align}
A_{\rm I} &\simeq 6.6\times 10^{-7} g^{7/2} \,,\label{eq:A1}\\
A_{\rm II} &\simeq 1.1 \times 10^{-10} g^{5/2} \ln\left(\frac{g}{0.0003}\right)^2\,. \label{eq:A2}
\end{align}
Therefore, $A_{\rm II}$ can be derived from $A_{\rm I}$
with the following expression:
\begin{equation}\label{eq:A2A1}
A_{\rm II} 
\simeq
(2.9\times 10^{-6})
A_{\rm I}^{5/7}
\left[ 
\ln A_{\rm I}^{4/7} + 24
\right]\,.
\end{equation}
The scale dependence of the contributions are given by the 
dimensionless functions
\begin{align}
f_1(x_i) &\equiv
\frac{\left[\sin(x_i)-{\rm SinIntegral}(x_i)\right]^2}{x_i^3} \,,\label{eq:f1}\\
f_2(x_i) &\equiv
\frac{-2x_i\cos(2x_i)+(1-x_i^2)\sin(2x_i)}{x_i^3}\,, \label{eq:f2}
\end{align}
where $x_i \equiv \frac{k}{k_i}$,
and $\text{SinIntegral}(x)=\int_0^{x}\frac{\sin z}{z}dz$.
The peaks of the functions $f_1(x)$ and $f_2(x)$
evaluate to $f_1^{\rm max} \simeq 0.11$ and $f_2^{\rm max}\simeq0.85$, respectively.
The parameter
%\footnote{The peak of dominant feature $f_1^{\rm max}$ occurs at $k \simeq 3.35 k_i$ and subdominant feature $f_2^{\rm max}$ occurs at $k \simeq 1.25k_i$.\label{fn:f1f2}} 
$k_i/{\rm Mpc}^{-1}$
is related to the number of e-folds
$N(\phi)$ from the end of inflation to 
the time of $i$-{th} burst of 
particle production
%\green
{as in eq.~\eqref{eq:Ne}.}
%\blue
{In fact, the peak of dominant feature $f_1^{\rm max}$ occurs at 
\begin{equation}
    k_p \simeq 3.35\times  k_i\,
    \label{eq:kpeak}
\end{equation}
and subdominant feature $f_2^{\rm max}$ occurs at $\simeq 1.25~k_i$
on the primordial power spectrum.}
%
%The location of each bump,

%The power spectrum with bump-like features
%given by eq.~\eqref{eq:PS_bump}
%is different from the templates investigated 
%in \cite{Akrami:2018odb}.
%The primordial power spectra from the
%base model and particle production 
%are plotted in figure~\ref{fig:Multibumps}
%for illustration.
%
%\magenta{I think Figure 1 is not needed.  Figs 2 and 3 illustrate the bump model. You can add the base model to them only.}

\begin{figure}[]
	\centering 
	\includegraphics[width=.48\textwidth]{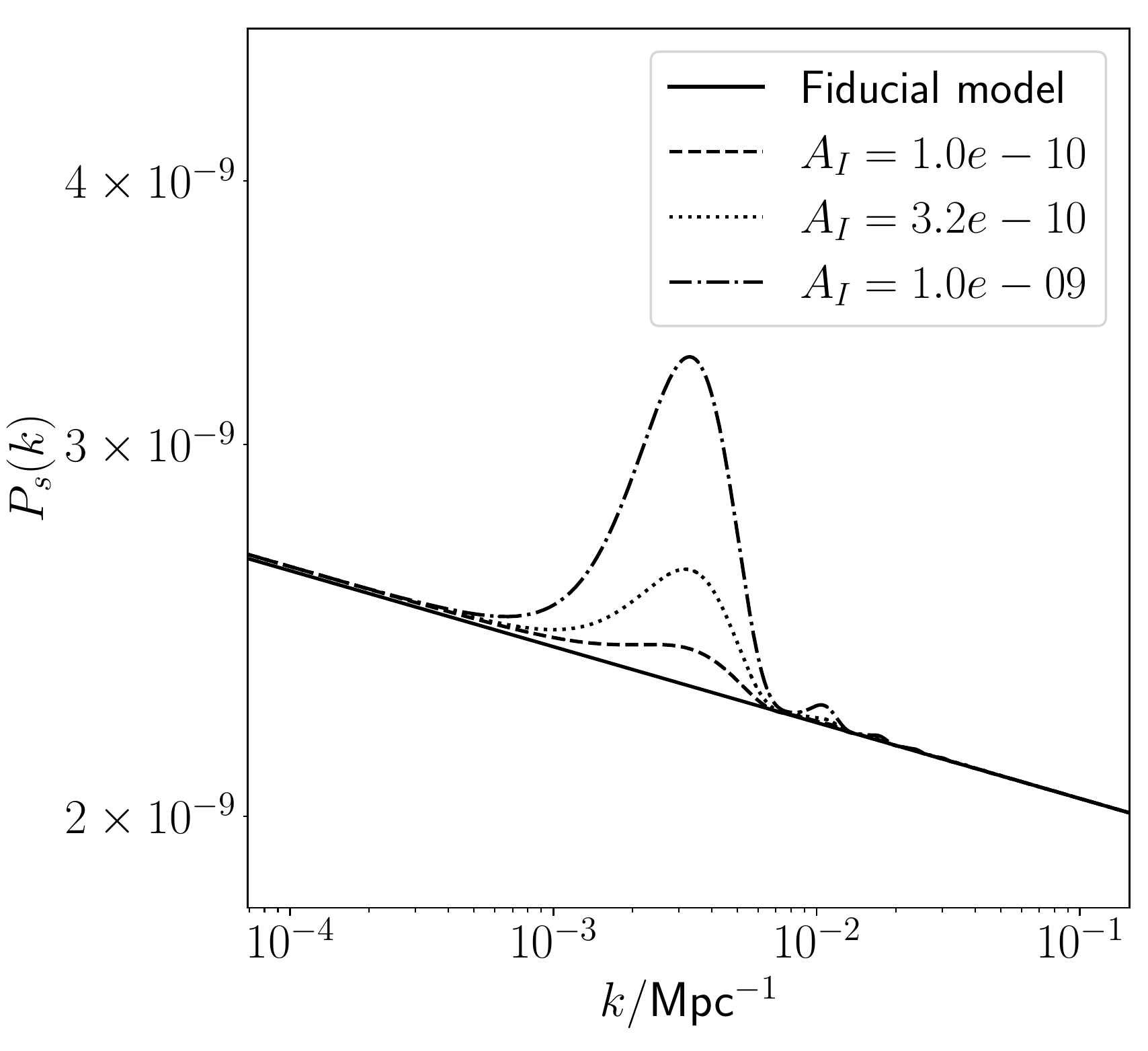}
	\hfill
	\includegraphics[width=.48\textwidth]{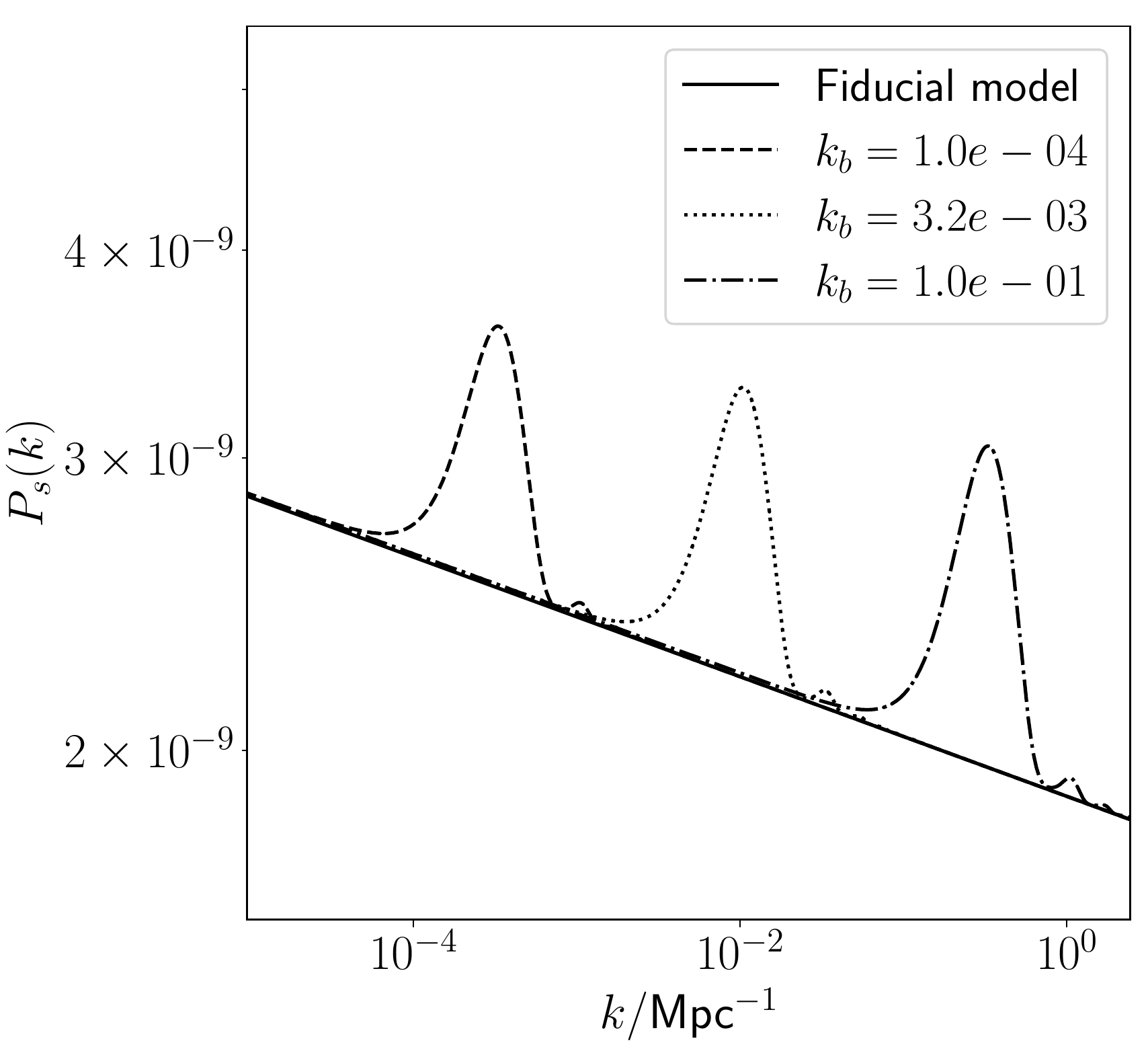}
	\caption{\label{fig:singlebump} \textit{Left}: Primordial power spectra for 
	single bump models for different $A_{\rm I}$ values, $k_b$ fixed at $1.0\times10^{-3}$ Mpc$^{-1}$;
	\textit{Right}: single bump models for different $k_b$ values
	(the peaks occur at $3.35\times k_b$), 
	$A_{\rm I}$ set at $1.0\times10^{-9}$.
	The primordial power spectra for the fiducial model is plotted
	in solid line.}
\end{figure}

%\begin{figure}[tbp]
%\centering 
%\includegraphics[width=.6\textwidth]{multiBumps_v1.pdf}
%\caption{\label{fig:Multibumps} The primordial scalar power spectra given by the 
%power law in eq.~\eqref{eq:PS_PowerLaw} (solid) and
%particle production in eq.~\eqref{eq:PS_bump} (dashed).
%The parameters chosen for plotting are: 
%$A_{\rm I}=5\times10^{-10},~k_1=1\times10^{-6} (\text{Mpc}^{-1}) %~\text{and}~\Delta=2$.}
%\end{figure}
%
%\magenta{$k_0$ should be $k_*$, right?}
%
%To understand the effect of each parameter 
%of our modified power spectrum,
%we consider a simplified model which gives rise to
%\magenta
{In this work, we use the form of the power spectrum given by eq.~\eqref{eq:PS_bump}}, which
is different from the templates investigated 
in \cite{Akrami:2018odb}.
We begin with a model which 
involves a single burst of particle production.
The primordial power spectrum for a
single bump having an amplitude $A_{\rm I}$ 
located at the scale $\sim 3.35\times k_b/\text{Mpc}^{-1}$
is given by
\begin{equation}\label{eq:single_bump}
P_s(k)
=
A_s \left(\frac{k}{k_\ast}\right)^{n_s-1}
+
A_{\rm I} \left(\frac{f_1(k/k_b)}{f_1^{\rm max}}\right)
+
A_{\rm II} \left(\frac{f_2(k/k_b)}{f_2^{\rm max}}\right)\,.
\end{equation}
The primordial power spectrum for different values of
$A_{\rm I}$ and $k_b$ are shown in figure~\ref{fig:singlebump}.
%The effects of parameters $A_{\rm I}$ and $k_b$
%on the primordial power spectrum 
%are shown in figure~\ref{fig:singlebump} for illustration.
Note that the parameter $k_b$ controls the width of the feature in addition to the feature's location via eq.~\eqref{eq:kpeak}.
%and width of the feature.}
%
%%%%%%%%%%%%%%%%%%%%%%%%%%%%%%%%%%%%%%%%%%%%
%\section{\sout{Data Analysis}} 
\section{Methodology}
\label{sec:method}
%\label{sec:dataAnalysis}
%%%%%%%%%%%%%%%%%%%%%%%%%%%%%%%%%%%%%%%%%%%%
%
We work in the framework of
Friedmann-Lema\^{i}tre-Robertson-Walker cosmology
with a flat spatial geometry. 
%We mainly consider two cosmological models
%in this analysis. 
%The first is the 
%%featureless 
%$\Lambda$CDM model,
%and the second is 
%our particle physics model of inflation
%that produces primordial features. 
%%
%In both the models, 
The $\Lambda$CDM model %cosmology
%contains
%the following:
%baryonic matter, cold dark matter,
%massive neutrinos, and
%constant dark energy.
%This is 
is parametrized by 
the baryon density 
$\omega_b = \Omega_b h^2$, 
the cold dark matter density 
$\omega_{\text{cdm}} = \Omega_{\text{cdm}} h^2$,
the present Hubble parameter $H_0$,
and the optical depth to reionization,
$\tau$.
%\noindent\green{(KF: it looks to me that this section
%is just a preparation for the analysis. The next section
%is the analysis. May consider combining with the next section.)}\\
%\magenta{(PC: I think we do need to separate the methodology from the results section. I have changed the title to 'Methodology'. We can move the first para of section 2 here. Table 2 for prior ranges overlaps with description in section 4. Mention that justification of the choice of values will be provided in section 4. )}

%\magenta
%{Below we briefly discuss the data products, the codes we use, and how we carry out the Bayesian inference.}
Below, we briefly explain how we carry out the Bayesian analysis, the data products, and the codes used for the analysis.

%===========================================
\subsection{Model selection}
%{Bayesian Analysis}
%\label{subsec:Bayesian}
\label{subsec:ModelSelec}
%===========================================
%
Let $M_0$  be the fiducial model with six parameters, and $M_1$ be the model that predicts primordial features
with $n$ additional parameters
(in this work, $n=2$ for the single bump model and
$3$ for the multi-bump model).
We compare the two models, $M_0$ and $M_1$, by the difference in their
log-likelihood at best-fit values.
Then, the quantity \textit{effective $\Delta \chi^2$}
is given by
\cite{Planck:2013jfk,Planck:2015sxf,Akrami:2018odb}:
\begin{equation}\label{eq:delchi2}
\Delta \chi^2_{\rm eff} := 
\chi^2_{M_1} 
-
\chi^2_{M_0}\,.
%	-2~(\ln \mathcal{L}_{max}(M_2) 
%	-  \ln \mathcal{L}_{max}(M_1))
\end{equation}
A negative value of \textit{$\Delta \chi^2_{\rm eff}$}
quantifies the improvement in the fit due 
to the addition of parameters. 
%Therefore, negative value of $\Delta \chi^2 = 
%\chi^2_{bump} 
%-
%\chi^2_{\Lambda CDM}$
%%	-2~(\ln \mathcal{L}_{\Lambda CDM}^{best fit} 
%%	-  \ln \mathcal{L}_{feature}^{best fit})$ can
%favor the model with feature over $\Lambda $CDM model.
For model comparison with $\geq 2$ degrees of freedom,
$|\Delta\chi^2_{\rm eff}|< 1$ may be considered a negligible improvement in the fit \cite{Press2007}. 
%as the corresponding confidence level is $\ll 68.3\%$ 
%for a model with $\geq 2$ degrees of freedom \cite{Press2007}.
%
\subsubsection*{Bayesian analysis}
In the Bayesian interpretation, 
probability measures a degree of belief.
%\paragraph{Bayes Theorem:} 
The Probability Distribution Function (PDF) of the model parameters
after obtaining the data $d$, 
i.e., posterior probability,
can be calculated using
the Bayes theorem:
\begin{equation}\label{eq:bayes}
p(\theta|d) = 
\frac{p(\theta)\, p(d|\theta)}
{p(d)}\,,
\end{equation}
where 
$p(\theta)$ is the prior probability of parameter $\theta$,
likelihood function $p(d|\theta)$ encloses experimental
measurements, 
and 
$p(d)=\int p(d|\theta)\, p(\theta)\, d\theta = \mathcal{Z}$ is called
the marginalized PDF of $d$, or Bayesian evidence. 

Calculation of evidence for a model plays a prime role
in the Bayesian model comparison.
The ratio of evidences of two competing models, $M_1$ and $M_0$,
is called the Bayes factor:
\begin{equation}
\mathcal{B}
=
\frac{\mathcal{Z}_{M_1}}{\mathcal{Z}_{M_0}}
\,.
\end{equation}
Interpretation of Bayes factor
was given by Jeffreys' scale \cite{Jeffreys:1939xee,Trotta:2005ar}
using the quantity $\ln \mathcal{B}$, 
and shown in table~\ref{table:jeffrey} for reference.
Negative values of $\ln \mathcal{B}$ mean
preference of model $M_0$ with respect to model $M_1$.\\[2mm]
 \renewcommand{\arraystretch}{1.3}
	\begin{table}
		\begin{center}
			\begin{tabular}{|c|p{7cm}|}
				\hline
				$\mathbf{\ln \mathcal{B}}$ & 
				\textbf{Preference of $M_1$ with respect to $M_0$}\\
				\hline
				0 - 1 & \hskip 2cm inconclusive \\
				1 - 2.5 & \hskip 2cm weak \\
				2.5 - 5 & \hskip 2cm moderate \\
				$>~5$ & \hskip 2cm strong \\
				\hline
			\end{tabular}
			\caption{Jeffreys' scale \cite{Jeffreys:1939xee} 
			for Bayesian model comparison, interpretations 
			modified by \cite{Trotta:2005ar}.}
			\label{table:jeffrey}
		\end{center}
	\end{table}
%
%\magenta
\noindent{\textbf{Data:}
%
%\textbf{Planck 2018}: 
We use a combination of the following data, %\magenta
{in the form of the angular power spectrum $C_{\ell}$}, from the \textit{Planck} 2018 data release~\cite{Planck:2018vyg}} 
(i) temperature and polarization data
covering multipoles $\ell \sim 30$ to $\ell \sim 2500$
(TE and EE covering up to $\ell \sim 1996$); 
%\textit{Planck} TT,TE,EE 
%\blue
{to avoid the exploration of the
high-$\ell$ nuisance parameters, 
we use compressed and faster likelihood
\texttt{Plik\_Lite}, 
which includes marginalization over foregrounds and residual systematics.}
%
%unbinned likelihood \texttt{Plik\_Lite-}TTTEEE
%(as we do not intend to vary the foreground nuisance %parameters, we use \texttt{Lite} version of the likelihood in which, nuisance parameter are marginalized),  
%
(ii) low $\ell$ polarization data (lowE)
covering $\ell =2 - 29$, 
%\textit{Planck} EE 
\texttt{Plik-}EE, 
(iii) low $\ell$ temperature data
covering $\ell =2 - 29$, 
%\textit{Planck} TT 
\texttt{Plik-}TT. \\[2mm]
\noindent \textbf{Codes:}
We use the Boltzmann code \texttt{CLASS}
\footnote{\url{https://github.com/lesgourg/class_public}} 
(The Cosmic Linear Anisotropy Solving System \cite{Lesgourgues:2011re,CLASSII:2011}) 
to calculate the angular power spectrum 
of the CMB anisotropies 
%$C_\ell$ defined in eq.~(\ref{C_l})
for a cosmological model.
To sample the parameter space, 
we use \texttt{MontePython}\footnote{\url{https://github.com/brinckmann/montepython_public}}
\cite{Brinckmann:2018cvx,Audren:2012wb}, 
which is a publicly available 
Markov Chain Monte Carlo (MCMC) sampling package in 
\texttt{Python}.
We use the nested sampling approach \cite{Skilling:2006gxv},
which calculates the Bayesian evidence for a model.
Nested sampling was incorporated in \texttt{MontePython}
by the sampling method \texttt{MultiNest} \cite{Feroz:2007kg,Feroz:2008xx}.
%\blue
{The triangle plots are generated with the package 
\texttt{GetDist}\cite{Lewis:2019xzd}.}
%
%%%%%%%%%%%%%%%%%%%%%%%%%%%%%%
\subsection{Selection of priors}
%%%%%%%%%%%%%%%%%%%%%%%%%%%%%%%%
%Bayesian analysis provides an elegant way to 
%incorporate our belief about the 
%model's parameters as a prior distribution.  
The selection of prior probability
is an essential ingredient in 
Bayesian statistics and reflects our 
initial knowledge about the model's parameters. 
Our selection of prior ranges of cosmological parameters 
%the $\Lambda$CDM model
is given in table~{\ref{table:priorCosmo}}.
The %\magenta{\sout{lower and upper bound} 
ranges of values of the 
fiducial model parameters
are taken to be a few times the \textit{Planck} $1\sigma$
values as provided in the recommended parameter file
by \texttt{MontePython} for multinest sampling.
We choose uniform probability distribution 
for the prior ranges of all the parameters.  
%considered in table~{\ref{table:priorCosmo}}.
The priors for the analysis
of single bump models are selected by taking different ranges of $\log k_b$ values. The justification for the different choices will be explained in the next section. 

%, as shown in figure~\ref{fig:priorkb}.
%By choosing different prior ranges of $\log k_b$,
%we compare our model with the CMB data at different 
%angular scales.
%The justification for the choice of our prior range 
%for the parameters of the bump model will be explained
%in the next section.
%%%%%%%%%%%%%%%%%%%%%%%%%%%%%%%%%
 \renewcommand{\arraystretch}{1.3}
\begin{table}
	\begin{center}
		\begin{tabular}{|p{1cm}|c|c|c|}
			\hline
			\multicolumn{2}{|c|}{\bf Model} & {\bf Parameters}
			& {\bf Prior ranges} \\
			\hline
			\multicolumn{2}{|c|}{\multirow{3}{*}{Common parameters for}}& $100~\Omega_b$ & [1.8, 3]\\
			\multicolumn{2}{|c|}{\multirow{5}{*}{and all other models}}
			& $\Omega_{\text{cdm}}$ & [0.1, 0.2] \\
			\multicolumn{2}{|c|}{fiducial model}& $h$ & [0.6, 0.8]\\
			\multicolumn{2}{|c|}{}& $\tau_\text{reio}$ & [0.004, 0.12] \\
			\multicolumn{2}{|c|}{}& $\ln 10^{10}A_{s }$ & [2.8904, 3.4012] \\
			\multicolumn{2}{|c|}{}& $n_s$ & [0.9, 1.1] \\
			\hline
			\parbox[t]{2mm}{\multirow{8}{*}{\hskip .3cm\rotatebox[origin=c]{90}{\bf Single bump models}}}
			& \multirow{2}{*}{Prior-A}& $\log A_{\rm I}$ & [$-13$, $-8$]\\
			&  & $\log k_b$ & [$-4.5$, $-1.0$] \\
			\cline{2-4}
			& \multirow{2}{*}{Prior-B} & $\log A_{\rm I}$ & [$-13$, $-8$]\\
			&  & $\log k_b$ & [$-4.5$, $-2.2$] \\
			\cline{2-4}
			& \multirow{2}{*}{Prior-C} & $\log A_{\rm I}$ & [$-13$, $-8$]\\
			& & $\log k_b$ & [$-2.2$, $-1.0$] \\
			\cline{2-4}
			& \multirow{2}{*}{Prior-D} & $\log A_{\rm I}$ & [$-13$, $-8$]\\
			& & $\log k_b$ & [$-3.2$, $-2$] \\
			\hline
			\multicolumn{2}{|c|}{\multirow{3}{*}{\bf Multi-bump model}} & $\log A_{\rm I}$ & [$-13$, $-8$]\\
			\multicolumn{2}{|c|}{}& $\log k_1$ & [$-4.5$, $-1.0$] \\
			\multicolumn{2}{|c|}{}& $\Delta$ & [$0$, $1$]\\
			\hline
		\end{tabular}
		\caption{The selection of prior ranges for the cosmological parameters of fiducial, single bump, and multi-bump models.}
		\label{table:priorCosmo}
	\end{center}
\end{table}
%
%
%\begin{figure}[tbp]
%	\centering 
%	\includegraphics[width=0.5\textwidth]{AllPrior_v3.png}
%	\caption{\label{fig:priorkb}
%    The prior ranges of the parameter $\log k_b$ selected for the analysis of single bump models.}
%\end{figure}
%
%%%%%%%%%%%%%%%%%%%%%%%%%%%%%%%%%%%%%%%%%%%%%%%%%%%%%%%%%%%%%%%%%%%%%%%
\section{Results}
\label{sec:results}
%%%%%%%%%%%%%%%%%%%%%%%%%%%%%%%%%%%%%%%%%%%
This section presents our results for the 
single bump and multi-bump models of particle production 
during inflation.
\subsection{Single burst of particle production}

We first carried out Bayesian analysis for
the fiducial model with 
six parameters:\\
{$~\Omega_{b },~\Omega_{cdm}$},$~h,~\tau,$ 
$~\ln 10^{10}A_{s },~n_s$, 
with the power spectrum given by eq.~\eqref{eq:PS_PowerLaw}.
We then incorporate a single bump 
{using the}
power spectrum given by eq.~\eqref{eq:single_bump} in \texttt{CLASS},
corresponding to a single burst of particle production during 
inflation and observable in CMB. 
This model has two additional parameters. 
Since the parameter $A_{\rm I}$ controls the amplitude of the bump, 
the upper limit of its prior range is chosen so that the posterior
probability becomes negligible well within the prior range. 
%There is no theoretical constraint on the lower bound of $A_{\rm I}$, so we choose it to be zero.
The lower bound on $A_{\rm I}$ %\magenta{\sout{could} can} 
can be derived from the
lower bound on model parameter $g$ using eq.~\eqref{eq:A1}.
The theoretically estimated lower bound given in \cite{Pearce:2017bdc} is $g^2 \gg 10^{-7}$,
which gives $A_{\rm I} > 10^{-17}$.
The increment to the CMB power spectrum 
due to a bump with such a small amplitude 
is of $\mathcal{O}(10^{-9})$, which is indistinguishable from 
the fiducial model.
Therefore, we choose the lower bound on $A_{\rm I}$
to be $10^{-13}$,
which contributes an $\mathcal{O}(10^{-5})$ increment 
in the CMB power spectrum.

Next, the parameter $k_b$ that controls the bump's location
is not restricted from the theoretical model. %\magenta
{In order to first understand its effect, we examine how varying $k_b$ affects 
the CMB power spectra,   $\mathcal{D}^{XX}_\ell \equiv \ell (\ell+1)C^{XX}_\ell/2\pi$, where $XX$ stands for either $TT$, $TE$ or $EE$. We obtain the residual $\Delta \mathcal{D}^{XX}_\ell$ as
\begin{equation}
\Delta \mathcal{D}^{XX}_\ell \equiv \Delta \mathcal{D}^{XX, {\rm bump}}_\ell - \Delta \mathcal{D}^{XX, {\rm fid}}_\ell,
\label{eq:DDl}
\end{equation}
where the superscripts `bump' and `fid' denote the bump and fiducial models, respectively.}

%%%%%%%
\begin{figure}[tbp]
	\centering 
	\includegraphics[width=\textwidth]{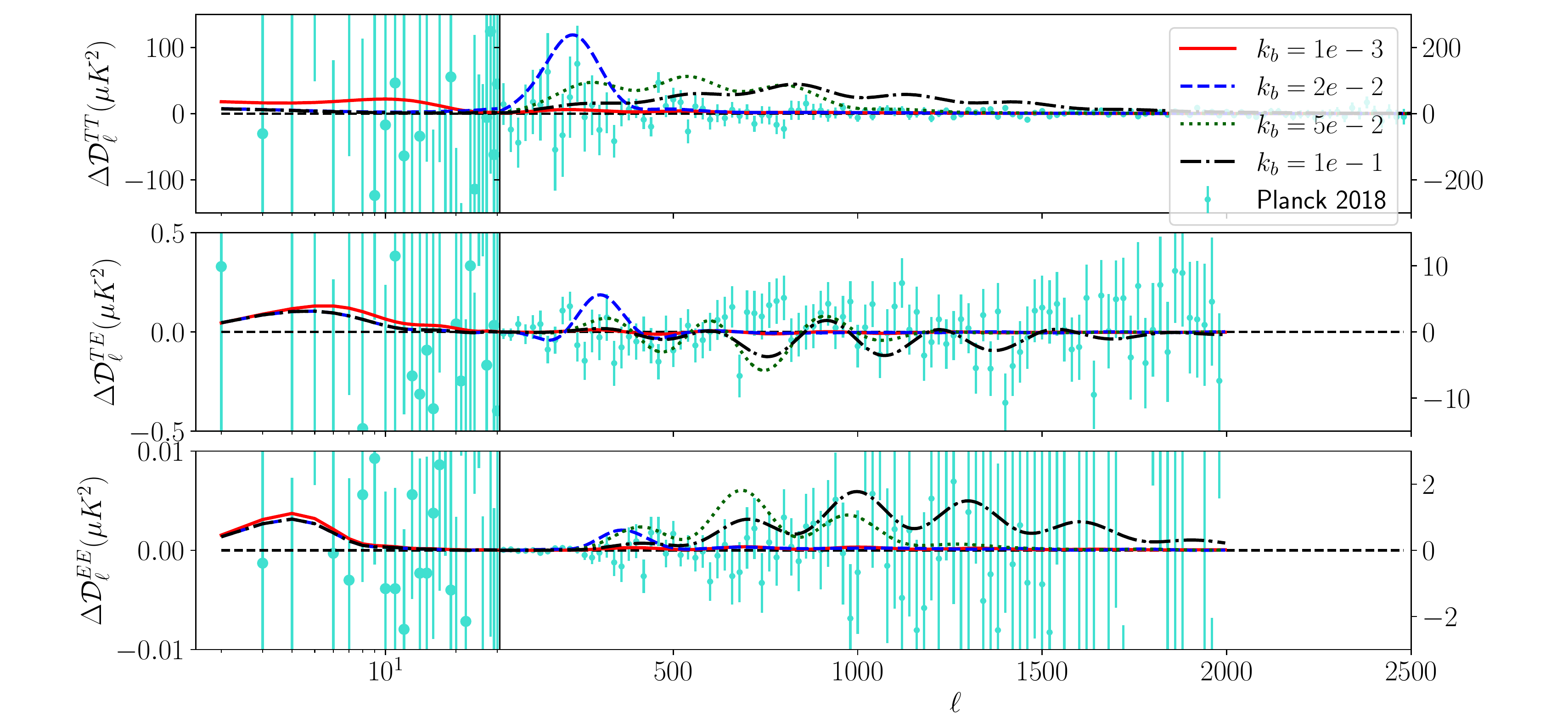}
	\caption{\label{fig:bump_residual}
	    The residual CMB  power spectra of single bump 
		models with respect to the fiducial model, %\magenta
		{defined by eq.~\eqref{eq:DDl}, for different values of $k_b$ and fixed $A_{\rm I}$}. 
		The three panels show CMB TT (top), TE (middle) and 
		EE (bottom) power spectra, respectively.
		The data points with error bars, %\magenta
		{shown in cyan,} are obtained 
		by subtracting the fiducial model from the
		observed CMB power spectrum (\textit{Planck} 2018). }
\end{figure}
%

%\magenta
{Figure~\ref{fig:bump_residual} shows $\Delta \mathcal{D}^{XX}_\ell$ for four different $k_b$ values, keeping $A_{\rm I}$ fixed to be $9\times10^{-10}$ for all the cases. We find that $\Delta \mathcal{D}^{TT}_\ell$ is positive at all $\ell$. It exhibits oscillations which consist of one primary peak at some scale $\ell$ roughly set by $k_b$, followed by peaks of decreasing heights on either side of the primary peak. $\Delta \mathcal{D}^{TE}_\ell$ and $\Delta \mathcal{D}^{EE}_\ell$ also exhibit roughly similar oscillations. $\Delta \mathcal{D}^{TE}_\ell$ takes both positive and negative values, while $\Delta \mathcal{D}^{EE}_\ell$ is positive at all $\ell$. Interestingly, we see an additional rise of power at $\ell < 10$ for TE and EE when a single bump is located at 
lower $k$ values. This can potentially lead to degeneracy with the optical depth to reionization, $\tau$. To examine this further, we compare the effect of the bump with the effect of varying $\tau$ in the angular power spectra. 
Figure~\ref{fig:bump_residual} shows that  incorporating primordial features at different $k_b$ affects the angular power spectrum at corresponding localized regions in $\ell$ space. 
It does not modify or shift the locations of the peaks of the oscillations. In comparison, varying $\tau$ 
introduces shifts of the peak locations, in addition to varying the height of the {\em reionization bump} for the polarization power spectrum at $\ell \le 10$
(see figure~\ref{fig:residual_diff_tau} in  appendix~\ref{app:degeneracy} for the CMB power spectra at different $\tau$).
Therefore, the degeneracy with the optical depth is broken when data from all the multipoles are considered.}
%
%Note also that the width of a bump \red{$<--$which bump are you talking about?} is controlled by the
%location parameter $k_b$ itself.
%As a result, a bump with higher $k_b$ appears
%broader in the CMB angular power spectrum.

%\magenta
{With the insight gained from the above study, we choose the following ranges of the prior for $k_b$: }%\blue{\sout{We compare the single bump models with the fiducial model using the following choices of prior ranges
%chosen with the help of the residual power spectra}:%\\[2mm]
%
\begin{itemize}
	\item Prior-A: $-4.5 < \log~k_b < -1.0$, a range of wave-numbers for $k_b$ such that the primordial bump-like features appear on all the multipole ranges 
	constrained by \textit{Planck}.
	%A range of wave-numbers for $k_b$ containing all the observable scales covered by the \textit{Planck}.
	%
	\item Prior-B: $-4.5 < \log~k_b < -2.2$, 
	where the bump incorporated in the primordial power spectrum appears only in the 
	lower multipole region of the CMB power spectrum.
	\item Prior C: $-2.2 < \log~k_b < -1.0$, where the contribution of the bump is 
	spread in the higher multipole region of the 
	CMB TT power spectrum.
	\item Prior D: $-3.2 < \log~k_b < -2.0$, a prior range in $k$-space 
	such that the primordial bump-like features
	peak around the intermediate scale of CMB data,
	or the multipole region $\ell \sim (30, 500)$.	
\end{itemize}

\begin{figure}[!tbp]
	\centering
	\begin{subfigure}{.6\textwidth}
	\includegraphics[width=\linewidth]{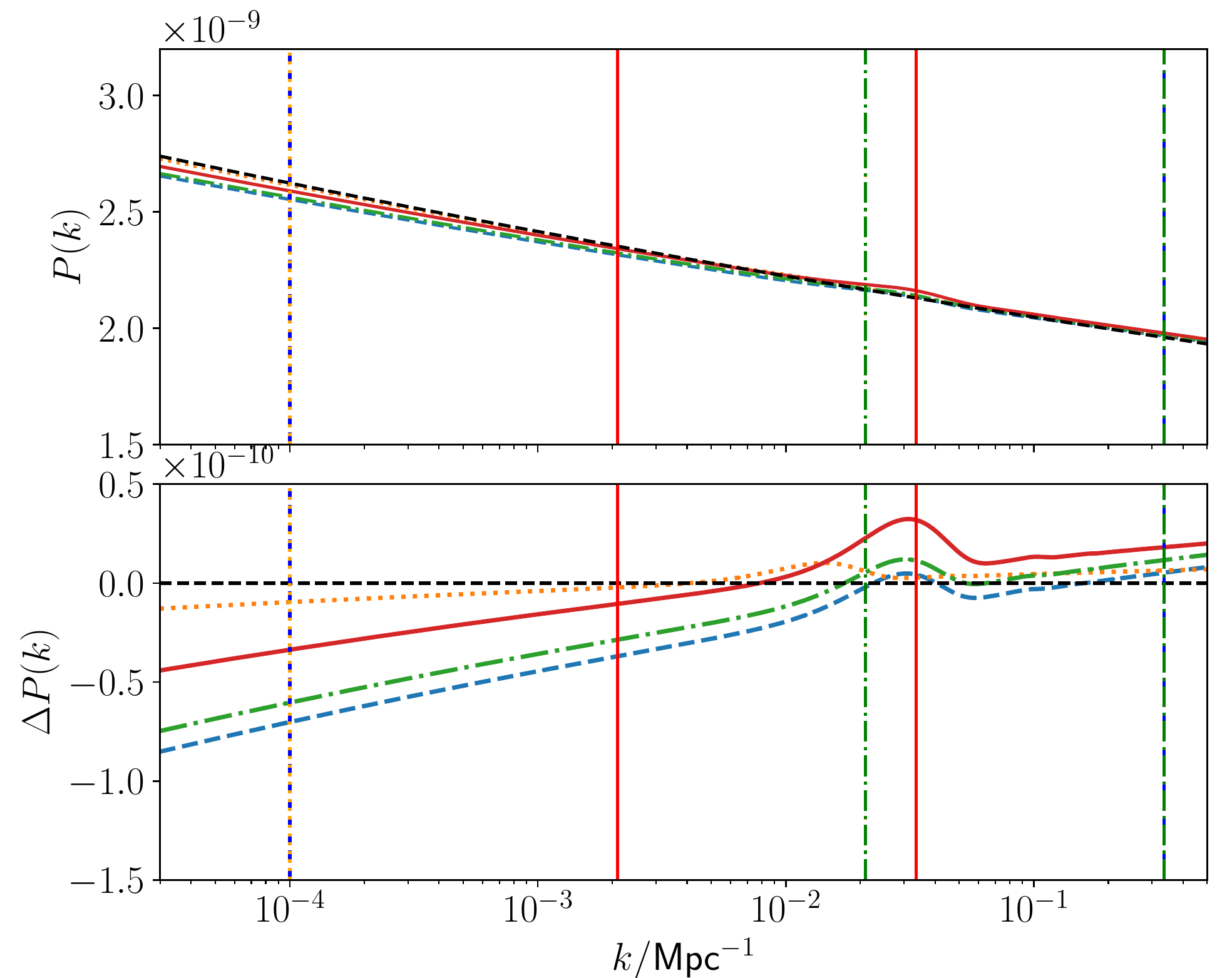}
	\end{subfigure}%
	\begin{subfigure}{.15\textwidth}
	    \includegraphics[width=\linewidth]{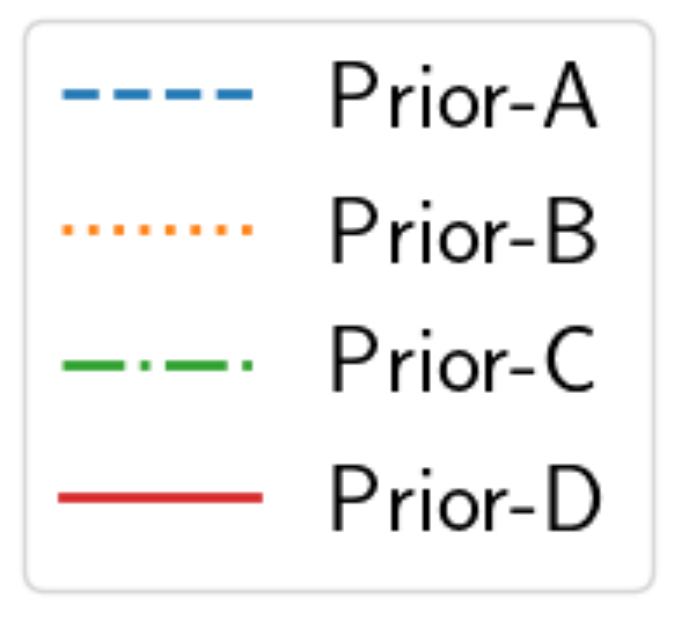}
	    %\label{fig:PPS_bf_singelBump}
	    \end{subfigure}\\
	\begin{subfigure}{0.8\textwidth}
	    \includegraphics[width=\linewidth]{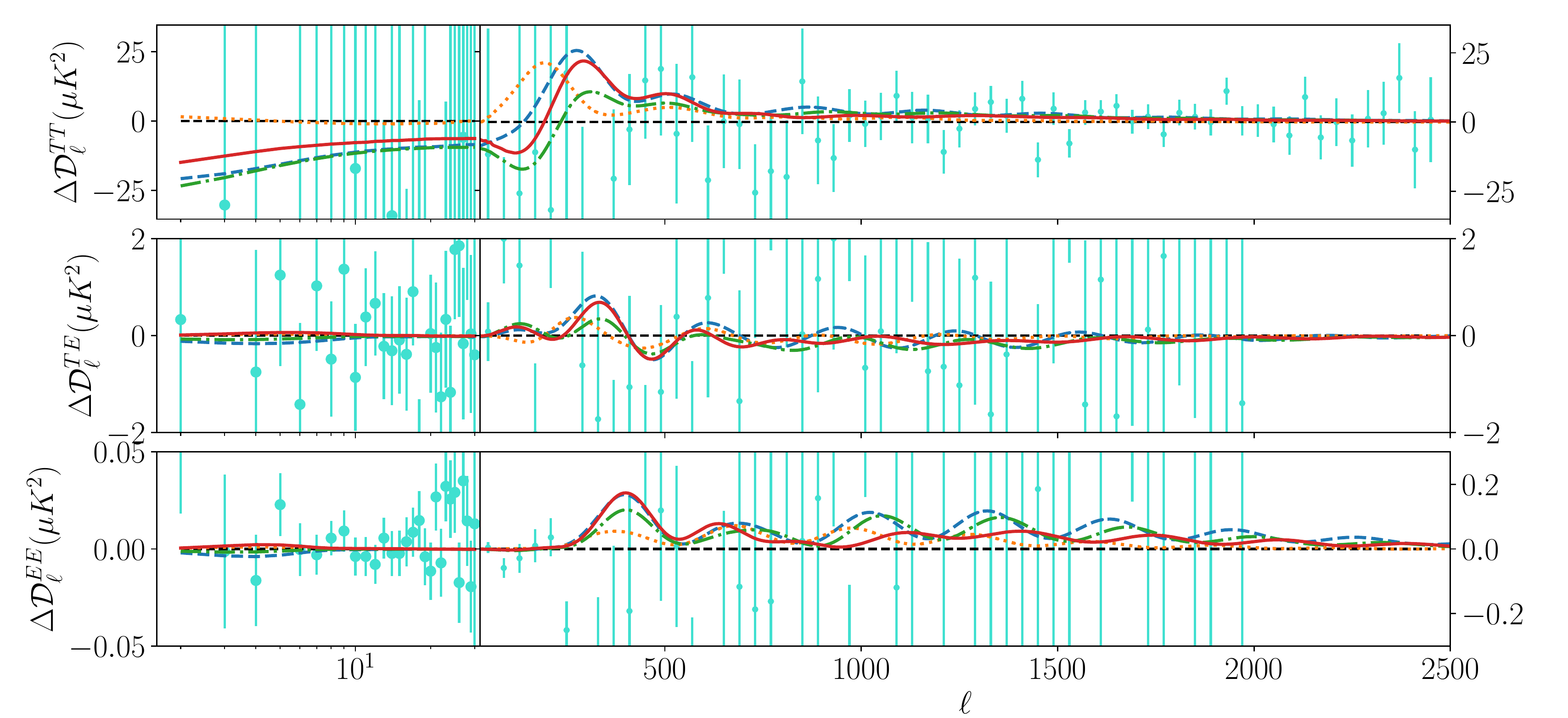}
	    %\label{fig:CMB_residual_singleBump}
	\end{subfigure}
	
	\caption{\label{fig:PPS_singleBump}{\it Top:} {The primordial power spectrum corresponding to best-fit parameters of the single bump models.
		{\it Middle:} Same as the top but difference in the power spectrum,
		$\Delta P(k):=P_{\rm bump}(k)-P_{\rm fiducial}(k)$, on the vertical axis. The peaks of the bumps are at $k_p \simeq 3.35~k_b$.
		%as mentioned in eq.~\eqref{eq:kpeak}. 
		The vertical lines mark our choices of prior range 
		in terms of $k_p$.
		{\it Bottom:} The residuals in CMB TT, TE, and EE 	power spectra of best-fit single bump models with respect to the fiducial model.
	The data points with error bars are obtained 
		by subtracting the fiducial model from the 
		observed CMB power spectrum by \textit{Planck} 2018.}
	}
\end{figure}

%\noindent \textbf{Prior-A}: In this prior, we choose 
%%\sout{started with} 
%a range of wave-numbers 
%{for $k_b$,}
%containing
%%\sout{covers} 
%all the observable scales covered by the \textit{Planck}.
%We call this prior range as \textbf{Prior-A}.
We use flat prior probability distribution in log-space for 
both $A_{\rm I}$ and $k_b$. 
The results of our analysis are shown in table~\ref{table:SingleBumpresult}. 
The first column shows all the parameters of the fiducial and single bump models. 
The following columns show the best-fit values and $95\%$ limits of these parameters.
%obtained for the prior-A.
%
A slight 
%\magenta
{relative shift in the values of the cosmological parameters} %between the fiducial and single bump models} 
can be noticed in the table. 
The shifts for all the parameters are within the 1$\sigma$ 
uncertainty levels obtained from the {\it Planck} results.
The last two rows are essential for model selection,
as discussed in section~\ref{subsec:ModelSelec}.
{We provide the improvement in fitting due to individual likelihoods for all the models we consider in 
table~\ref{tab:chi2_separate}. 
%\blue{Table~\ref{tab:chi2_separate} shows 
%the improvement in fitting due to individual likelihoods for all the models we consider.  
%We plot one and two-dimensional 
%posterior distributions for only the power spectrum parameters in figure~\ref{fig:bump_priorB_C}.
%The entire triangle plots for all the parameters of
%the models are given in appendix~\ref{app:2dpdf}. 
We also plot the best-fit power spectrum for different choices of priors in the top and middle panels of figure~\ref{fig:PPS_singleBump}.
The bottom panel shows corresponding CMB residual power spectra.}

\begin{figure}[tbp]
	\centering 
	\begin{subfigure}{.5\textwidth}
		\includegraphics[width=\linewidth]{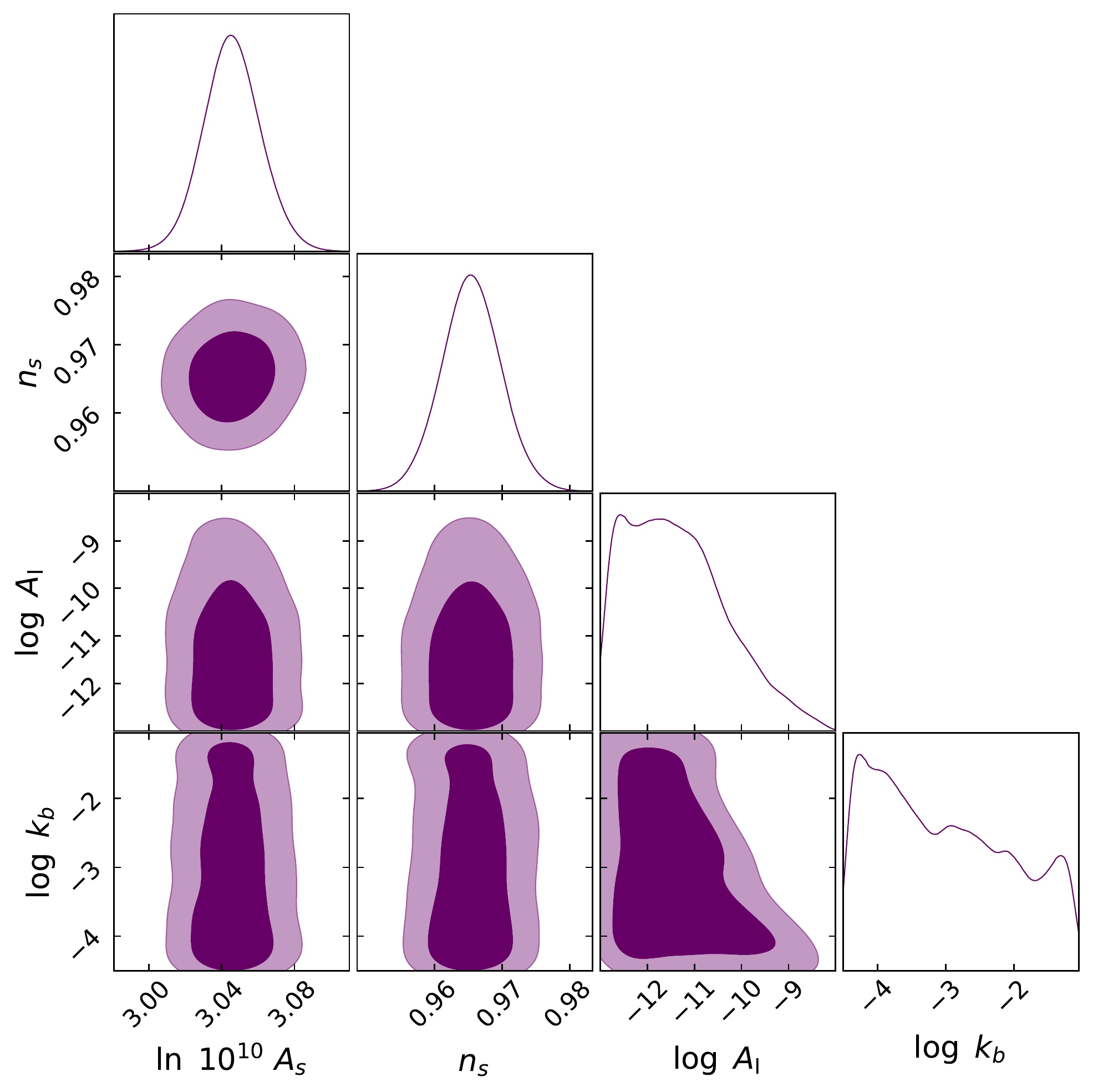}
		\caption{Prior-A}
	\end{subfigure}%
	\begin{subfigure}{.5\textwidth}
		\includegraphics[width=\linewidth]{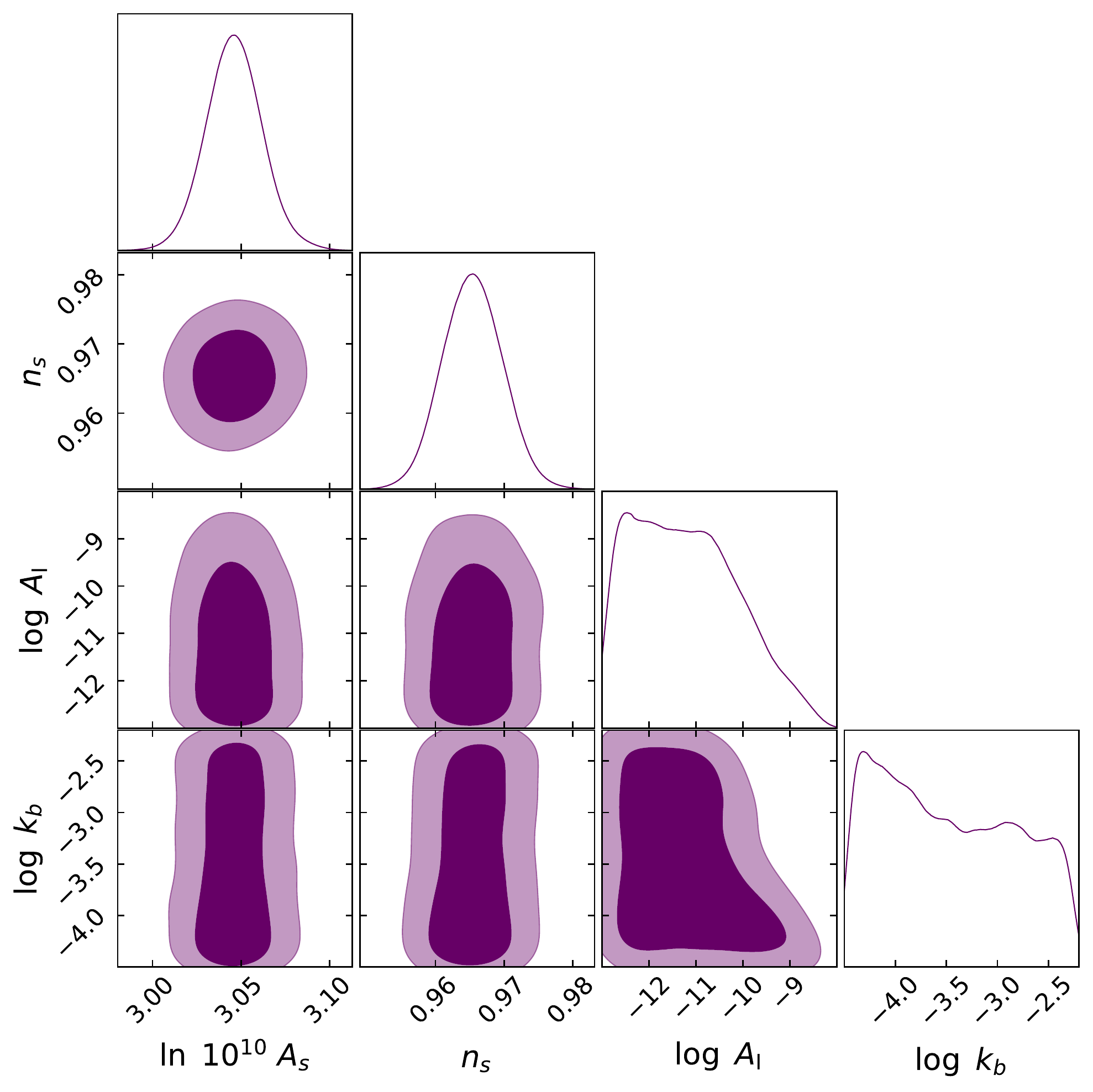}
	\caption{Prior-B}
	\end{subfigure}\\
	\begin{subfigure}{.5\textwidth}
		\includegraphics[width=\linewidth]{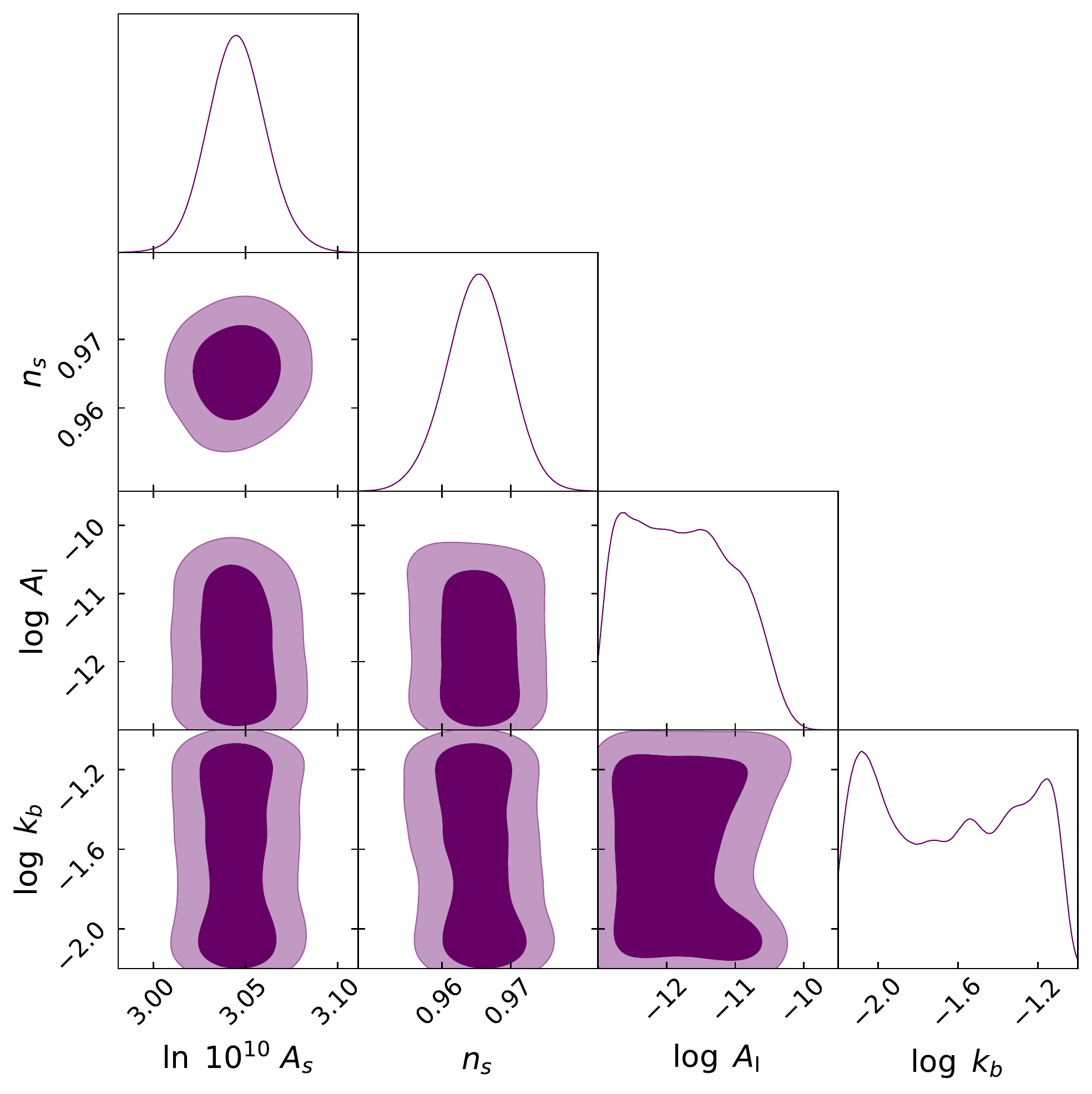}
	\caption{Prior-C}
	\end{subfigure}%
	\begin{subfigure}{.5\textwidth}
		\includegraphics[width=\linewidth]{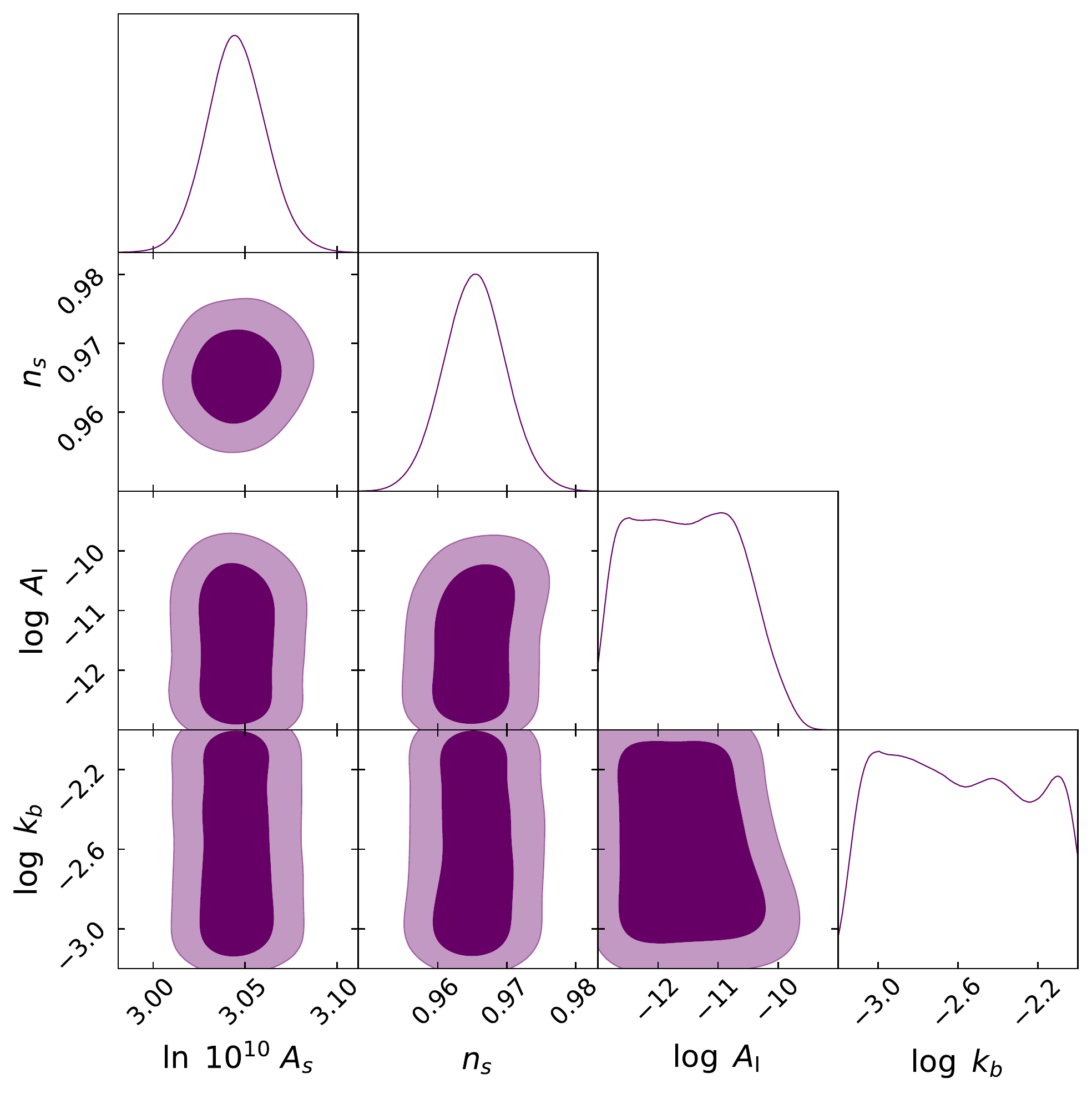}
	\caption{Prior-D}
	\end{subfigure}	
    \caption{\label{fig:bump_priorB_C} {The one and two-dimensional posterior distributions of 
	the parameters 
	%\magenta
	{for the single bump model corresponding to the} power spectrum
		eq.~\eqref{eq:single_bump} for different choices of prior. 
		The light and dark shaded regions indicate parameter space 
		corresponding to 68\% and 95\% confidence levels, respectively.}
	}
\end{figure}

\renewcommand{\arraystretch}{1.7}	
\begin{table}[!tbp]
	\centering
		\resizebox{\textwidth}{!}
				{\begin{tabular}{|l|c|c|c|c|c|c|c|c|c|c|}
				\hline
				\multirow{2}{*}{\bf Parameters} & \multicolumn{2}{c|}{\multirow{2}{*}{\bf Fiducial model}}
				& \multicolumn{8}{c|}{\bf Single-bump model with different priors}\\
				\cline{4-11} 
				& \multicolumn{2}{c|}{} & \multicolumn{2}{c|}{\bf Prior-A} & \multicolumn{2}{c|}{\bf Prior-B }& 
				\multicolumn{2}{c|}{\bf Prior-C }& 
				\multicolumn{2}{c|}{\bf Prior-D} \\
				\cline{2-11}
				& Best-fit & $95\%$ limits & 
				Best-fit & $95\%$ limits &
				Best-fit & $95\%$ limits &
				Best-fit & $95\%$ limits &
				Best-fit & $95\%$ limits \\
				%\cline{3-7}
				%& & $\ell \sim 467$ & $\ell \sim 950$ & $\ell \sim   \\
				%
				\hline
				${100~\Omega_{b }}$ &
				$2.237$ &
				$2.236^{+0.030}_{-0.029}$&
				$2.239$ &
				 $2.236^{+0.029}_{-0.028}$&
				$2.235$&
				$2.236^{+0.028}_{-0.029}$&
				$2.231$ &
				$2.236^{+0.029}_{-0.029}$&
				$2.233$ &
				$2.234^{+0.030}_{-0.030} $\\

				${\Omega_{cdm }}$ & 
				$0.1206$& 
				$0.1203^{+0.0027}_{-0.0027}$&
				$0.1204$& 
				$0.1202^{+0.0027}_{-0.0026}$&
				$0.1202$& 
				$0.1203^{+0.0027}_{-0.0027}$&
				$0.1211$ &
				$0.1203^{+0.0027}_{-0.0026}$&
				$0.1213$ &
				$0.1202^{+0.0027}_{-0.0027}$\\
				${h}$ &
				$0.6716$ & 
				$0.673^{+0.012}_{-0.012}  $ &
				$0.6724$& 
				$0.673^{+0.012}_{-0.012}$&
				$0.6731$& 
				 $0.673^{+0.012}_{-0.012}$&
				$0.6692$ &
				$0.673^{+0.012}_{-0.012}$&
				$0.6688$ &
				$0.673^{+0.012}_{-0.012} $\\
				${\tau}$ &
				$0.05357$ & 
				$0.055^{+0.016}_{-0.015}   $&
				$0.05127$& 
				$0.055^{+0.015}_{-0.015} $&
				$0.05438$& 
				 $0.055^{+0.016}_{-0.015}$&
				$0.05287$ &
				$0.055^{+0.016}_{-0.015}$&
				$0.05265$ &
				 $0.055^{+0.016}_{-0.015}$\\
				${\ln 10^{10}A_{s }}$ &
				$3.044$& 
				$3.046^{+0.033}_{-0.031}$ &
				$3.039$& 
				$3.046^{+0.032}_{-0.031}$&
				$3.046$& 
				$3.046^{+0.032}_{-0.031}$&
				$3.043$ &
				$3.045^{+0.033}_{-0.031}$&
				$3.048$ &
				$3.045^{+0.033}_{-0.031}$\\
				${n_{s }}$ & 
				$0.9641$ & 
				$0.9653^{+0.0086}_{-0.0084}$&
				$0.9678$& 
				$0.9655^{+0.0089}_{-0.0087}$&
				$0.9650$& 
				 $0.9654^{+0.0086}_{-0.0083}$&
				$0.9678$ &
				$0.9653^{+0.0086}_{-0.0092}$&
				$0.9669$ &
				$0.9654^{+0.0090}_{-0.0089}$\\
				%
				%				$\log A_{\rm I}$ &
				%				$-$ &
				%				$-10.86$ & 
				%				$-11.68$  & 
				%				$-10.96$ \\
				%
				$\log~{A_{\rm I}}$ &
				$-$ & 
				$-$ &
				%$1.85 \times 10^{-11}$ &
				$-10.73$ &
				$-11.4^{+2.1}_{-1.6}$&
				%$9.17 \times 10^{-12}$ &
				$-11.04$ &
				$-11.2^{+2.1}_{-1.8}$&
				%$1.83 \times 10^{-11}$ &
				$-10.74$ &
				 $-11.8^{+1.3}_{-1.2}$&
				%$2.7 \times 10^{-11}$ &
				$-10.57$ &
				 $-11.6^{+1.5}_{-1.4} $\\
				%
				%${k_b/\text{Mpc}^{-1}}$ & 
				$\log~k_b$ &
				$-$ & $-$ &
				%$0.0086$& 
				$-2.07$ &
				$-3.0^{+1.9}_{-1.5}$
				 &
				%$0.0042$&  
				$-2.38$ &
				$-3.48^{+1.1}_{-0.99}$ &
				%$0.0086$ &
				$-2.07$ &
				$-1.64^{+0.56}_{-0.55}$&
				%$0.0091$ &
				$-2.04$ &
				$-2.59^{+0.59}_{-0.55}$\\
				%
				%				$-\ln{\cal L}_\mathrm{min}$ & 
				%				$502.209$ &  
				%				$502.005$ & 
				%				$502.332$ & 
				%				$501.897$\\ 
				%				%
				%				minimum $\chi^2$ & 
				%				$1004$ & 
				%				$1004$ & 
				%				$1005$& 
				%				$1004$\\
				\cline{2-11}
				${\Delta \chi^2_{\rm eff}}$ & \multicolumn{2}{c|}{$-$} &
				\multicolumn{2}{c|}{$-0.29$} & 
				\multicolumn{2}{c|}{$-0.18$} & 
				\multicolumn{2}{c|}{$-0.96$}  & 
				\multicolumn{2}{c|}{$-1.12$}\\
				%
				%				NS log-evidence & 
				%				$-526.3204 \pm 0.0719685$ & 
				%				$-525.870 \pm 0.142224$& 
				%				$-526.096 \pm 0.142862$ & 
				%				$-526.226 \pm  0.143386$ \\
				%
				${\ln \mathcal{B}}$ & 
				\multicolumn{2}{c|}{$-$} & 
				\multicolumn{2}{c|}{0.4} &
				\multicolumn{2}{c|}{0.3} &
				\multicolumn{2}{c|}{0.2} &
				\multicolumn{2}{c|}{0.4} \\
				\hline
				%\bottomrule
		\end{tabular}}
	\caption{\label{table:SingleBumpresult} {Best-fit and {$95\%$ limits} of parameters for the fiducial model and the single-bump model for different choices of the prior range}.
	$\Delta \chi^2_{\rm eff}$ and the Bayes factor $\ln \mathcal{B}$ are quoted in the last two rows.}
\end{table}
    \renewcommand{\arraystretch}{1.3}
\begin{table}[h]
	\centering
	\begin{tabular}{|c|c|c|c|c|c|}
		\hline
		\multicolumn{2}{|c}{\multirow{2}{*}{\bf Model}}& \multicolumn{4}{|c|}{$\mathbf{\Delta \chi^2_{\rm eff}}$} \\
		\cline{3-6}
		\multicolumn{2}{|c|}{} & TTTEEE (high $\ell$) & TT (low $\ell$) & EE (low $\ell$) & {\bf Total}\\
		\hline
		& Prior-A & $1.06$ & $-1.04$ & $-0.30$ & $-0.29$ \\
		{\bf Single bump} & Prior-B & $-0.13$ & $-0.16$ & $0.10$ & $-0.18$ \\
		{\bf models} & Prior-C & $0.16$ & $-0.96$ & $-0.16$ & $-0.96$ \\
		& Prior-D & $-0.58$ & $-0.73$ & $0.19$ & $-1.12$ \\
		\hline
		\multicolumn{2}{|c|}{\bf Multi-bump model} & $-2.22$ & $-1.40$ & $0.55$ & $-3.07$ \\
		\hline
	\end{tabular}
	\caption{$\Delta \chi^2_{\rm eff}$ of the single bump and multi-bump models calculated for individual likelihoods.}
	\label{tab:chi2_separate}
\end{table}
\renewcommand{\arraystretch}{1}
%
%\paragraph{Prior A:}
%
The best-fit values and {$95\%$ limits} of all the parameters 
obtained from the Bayesian analysis with {\bf Prior-A} are given 
in the fourth and fifth columns of table~\ref{table:SingleBumpresult}.
We notice that the value of $\chi^2$ has not changed 
much by adding two more parameters.
Referring to Jeffrey's scale in table~\ref{table:jeffrey},
the Bayes factor $|\ln \mathcal{B}| < 1$
indicates an inconclusive preference toward 
the models under consideration.
The one and two-dimensional marginalized posterior distributions of the power spectrum parameters 
for Prior-A are shown on the top-left side of
figure~\ref{fig:bump_priorB_C}.

%Next, we compare our model with the CMB data by choosing 
%different prior ranges for the parameter $k_b$.
%As noticed from figure~\ref{fig:bump_residual}, 
%a bump takes multiple oscillatory shapes 
%with broad width when situated beyond a specific scale $k$. 
%Therefore, we consider the following two prior ranges
%for the subsequent analyses:
%(i) Prior B - $3 \times 10^{-5} < k/\text{Mpc}^{-1} < 6\times 10^{-3}$, 
%where bump incorporated in the primordial power spectrum 
%affects locally in the 
%lower multipole region of the CMB power spectrum
%and (ii) Prior C - $6\times 10^{-3} < k/\text{Mpc}^{-1} < 0.1$,
%where the contribution of the bump is 
%spread in the higher multipole region of the 
%CMB TT power spectrum.
%Next, we carry out Bayesian analysis with
%these two prior ranges.\\[2mm]
%
%\noindent \textbf{Prior B:}
The best-fit values and $95\%$ limits of the parameters for the single bump model 
with \textbf{Prior-B} (primordial features appearing at the lower $\ell$ of the CMB power spectrum) are given in the sixth and seventh columns 
of table~\ref{table:SingleBumpresult}.
No significant change in $\chi^2$ value was noticed 
due to the primordial feature
in this region. 
The value of the Bayes factor is $|\ln \mathcal{B}| < 1$
%and referring to the Jeffreys' scale %(table~\ref{table:jeffrey}) 
implies an inconclusive preference of models under consideration.
The one and two-dimensional marginalized posterior distributions of the power spectrum parameters 
for Prior-B are shown on the top-right side of
figure~\ref{fig:bump_priorB_C}.

Next, we estimate 
the maximum amplitude of primordial features that 
the CMB low-$\ell$ data may accommodate.  
The $95\%$ %\green
{Confidence Level} (CL)
upper limit on $A_{\rm I}$ obtained from one-dimensional
marginalized posterior distribution
is $\sim 7 \times 10^{-10}$, or about $35\%$ of 
the amplitude of scalar perturbations in the fiducial model, $A_s$. 
%The amplitude of subdominant correction $A_{\rm II}$
%depends on $A_{\rm I}$ via eq.~\eqref{eq:A1} and 
%eq.~\eqref{eq:A2}. 
%Therefore, the $95\%$ CL upper bound on $A_{\rm II}$ is 
%$\sim 1 \times 10^{-11}$ or
%$0.5\% A_s$.
The amplitude of the observable features helps to put 
a constraint on one of the theoretical model parameters. 
From eq.~\eqref{eq:A1}, 
we found an upper limit on the dimensionless coupling constant, $g$,
in the region chosen for Prior-B:
\begin{equation}
g < 0.1\,.
\label{eq:PAg}
\end{equation}
This constraint on the parameter $g$
leaves a large region of natural values.\footnote{%
In Quantum Field Theory (QFT),
%${\cal O}(1)$ 
a value of order one
is generically regarded as
natural magnitude for a dimensionless 
coupling constant
in the standard normalization.
Note that naturalness 
is not a very strict requirement,
and a value smaller by 
one order or even two may be tolerated.
Furthermore, if symmetry is restored in the limit
a coupling constant goes to zero,
a value much smaller than one 
can also be regarded
as natural for this coupling constant \cite{tHooft:1979rat}.
For the models based on 
higher dimensional gauge theory (eq.~\eqref{eq:S4}),
the gauge field becomes free
in the limit the parameter $g$ goes to zero,
and symmetry recovers \cite{Furuuchi:2014cwa}.
Therefore, values of $g$ much smaller than one 
are natural in this class of models.\label{fn:g}}
% There is a conjecture 
% (Weak Gravity Conjecture, WGC)
% which proposes that
% a gauge coupling constant much smaller than one
% could be in tension with consistency with quantum gravity.
% See \cite{Furuuchi:2020klq,Furuuchi:2020ery}
% for the constraints from WGC in explicit models
%of particle productions during inflation.}
%
%\green{implications to the theoretical models to be added.
%The same for other priors.}\\[2mm]
%

%

%\noindent \textbf{Prior C:}
The search for bump-like features on the higher multipoles of the CMB data with {\bf Prior-C} gives 
best-fit  values and $95\%$ limits, 
as quoted in the eighth and ninth columns 
of table~\ref{table:SingleBumpresult}.
The best-fit feature is
located around $9\times 10^{-3} \text{Mpc}^{-1}$,
giving an improved $\Delta \chi^2_{\rm eff} \sim -1$.  
{From table~\ref{tab:chi2_separate}, it can be seen that the low-$\ell$ temperature data contributes more to improve the fitting.}
The value of the Bayes factor again indicates
an inconclusive preference for models with available data.
The one and two-dimensional marginalized posterior distributions of the power spectrum parameters are shown on the bottom-left side of
figure~\ref{fig:bump_priorB_C}.
From one dimensional marginalized PDF, 
the upper bound ($95\%$ CL) on $A_{\rm I}$ 
%and $A_{\rm II}$
was found to be $\sim 3 \times 10^{-11}$,
or $1.5\%A_s$. 
%and $\sim 9 \times 10^{-13}$
%($0.05\%A_s$), respectively.
From eq.~\eqref{eq:A1},
we found that
\begin{equation}
g < 6 \times 10^{-2} \,.
\end{equation}

%
%\begin{figure}[tbp]
%	\centering 
%	\includegraphics[width=0.7\textwidth]{bump_k_gtr_1e-2_logPriorkb_triangle.pdf}
%	\caption{\label{fig:bump_priorC} The 1D and 2D posterior distributions 
%	of the parameters in power spectrum
%		eq.~\eqref{eq:single_bump} for the prior C. 
%		The light and dark shaded regions indicate parameter space 
%		corresponding to 68\% and 95\% confidence level respectively.
%	}
%\end{figure}
%

\noindent \textbf{Investigation of localized features:}
To avoid the multipole region where 
the primordial features are spread out,
next, we carry out Bayesian analysis with \textbf{Prior-D}.
%choose a prior range in $k$-space 
%such that the primordial bump-like features
%peak around the intermediate scale of CMB data,
%or the multipole region $\ell \sim (30, 500)$.
%%incorporates bumps in the multipole region 
%%$\ell \sim (30, 500)$ of the CMB power spectrum.
%We refer to this prior range as \textbf{Prior-D} -
%$6.3\times10^{-4} < k/\text{Mpc}^{-1} < 1.0\times10^{-2} $.
Results of the analysis with Prior-D are given in 
the last two columns of table~\ref{table:SingleBumpresult}.
{Interestingly, we find that the best-fit value
of the parameter $k_b$ points closer
to the one obtained in cases of Prior-A and C,
giving a slight improvement in the $\chi^2$ value. 
The corresponding primordial power spectrum and CMB residual power spectrum in figure~\ref{fig:PPS_singleBump}
may help the reader visualize the same.
%\footnote{Note that, as explained in footnote~\ref{fn:f1f2}, the best-fit location of the bump is at $k\simeq 3.35k_b \simeq 0.03 \text{Mpc}^{-1}$ 
%for the case of Prior-A, C and D.}. 
}
%
%In this case, also, we noticed a slight improvement 
%in the $\chi^2$ value 
%with $\Delta \chi^2_{\rm eff} \sim -1$.
%However, the Bayes factor %$\ln \mathcal{B}_{ij}$ was 
%less than one.
%showed an inconclusive preference (table~\ref{table:jeffrey}).
The one and two-dimensional marginalized posterior distributions of the power spectrum parameters are shown on the bottom-right side of
figure~\ref{fig:bump_priorB_C}.
The $95\%$ 
%\green
CL upper limit on $A_{\rm I}$ % and $A_{\rm II}$ 
was found to be $\sim 8 \times 10^{-11}$, or
$4\%A_s$.
%and $\sim 2 \times 10^{-12}$ ($0.09\%A_s$), respectively.
From eq.~\eqref{eq:A1}, we found that
\begin{equation}
g < 8 \times 10^{-2} \,.
\end{equation}

With the motivation to explain a slight excess in the CMB residual power spectra,
%near $\ell: 10-20$ and $\ell: 200-500$, 
we have also carried out Bayesian analyses of single bump
models with narrow prior ranges and uniform probability 
distribution in $k_b$. 
Including a narrow prior in $k_b$ that incorporates features 
near $\ell: 200-500$, we obtained 
$\Delta \chi^2_{\rm eff} = -1.1$
with the best-fit value $k_b$ same as 
Priors A, C and D, confirming our previous results.
%and best-fit feature located at 
%$9\times 10^{-3} \text{Mpc}^{-1}$
%(similar to the case of Prior-A, C and D).
Bayesian analysis of another narrow prior in $k_b$
that incorporates features near $\ell: 10-20$
showed a negligible improvement in the fit. 
We observed no significant improvement in the 
Bayes factor values for both cases.

{We remark that our results of the single bump models show prior dependence as the different prior ranges of $k_b$ search for features on different angular scales of the CMB data. We also note that the parameter $k_b$ is not constrained for any of the analysed
models, as shown in figure~\ref{fig:bump_priorB_C}. 
However, the analyses with multiple prior choices prefer a single bump at almost the same location (see figure~\ref{fig:PPS_singleBump}). 
We also obtain upper bounds on the amplitudes of primordial features predicted from particle physics models of inflation on different scales probed by the \textit{Planck}.}

\subsection{Multiple bursts of particle production}
As explained in section~\ref{sec:model}, multiple
bursts of particle production during inflation
may produce multiple bump-like features on the
primordial power spectrum \cite{Barnaby:2009dd,Pearce:2017bdc,Barnaby:2009mc,Chung:1999ve}.
The location of the $i$-{th} bump is parameterized 
%\magenta
{by $k_i = e^{(i-1)\Delta} k_1$, where 
$k_1$ specifies the location of the first bump via eq.~\eqref{eq:kpeak} while $\Delta$ 
controlling the spacing between subsequent bumps}. 
Here, we take $\Delta$ to be constant, 
as its $\phi$-dependence does not significantly change the 
qualitative feature of the power spectrum
in the parameter region we will consider. 
%\magenta
{Therefore, we have three additional parameters in the multi-bump model compared to the fiducial model: $A_{\rm I}$, $k_1$, and $\Delta$.}
%$\Delta$and given by $k_i = e^{(i-1)\Delta} k_1$.
%
%The models of inflation reviewed in Sec.~\ref{sec:model}
%predict continuous bump-like features on the 
%primordial power spectrum.

%\magenta
{We choose the same prior range for the parameter $A_{\rm I}$ as the single bump model. We need to specify the prior ranges for $k_1$ and $\Delta$.}  
%\sout{features} 
The number of bumps could be large enough
to cover the whole observable range.
We consider a phenomenological model
in which the first episode of a burst of particle
production can occur at any scale $k_1$, 
%\sout{and then continue to produce} 
followed by a  series of 
%\sout{features}
bumps over the 
scales $k > k_1$.
Therefore, we allow the prior range of $k_1$ to be as broad as 
the observable scale of \textit{Planck}. 
%\magenta
{Next, we need to specify a prior range for  $\Delta$.} 
%\magenta
{Note that} if the features are widely spaced,
% ($\Delta > 1$), 
%\magenta
{we expect that} the constraints on each bump will be %\magenta
{similar to} 
that of single bump models. 
When $0 \le \Delta \le 1$,
the multi-bumps on 
the primordial power spectrum 
largely overlap. 
%\magenta
{In what follows, we present the result of Bayesian analysis for the multi-bump model for this choice of the prior range of $\Delta$ and describe how it compares with the fiducial model.}
%to find the 
%signatures of multi-bumps in the \textit{Planck} data. 
%
\renewcommand{\arraystretch}{1.3}
\begin{table}[!t]
	\centering
		\begin{tabular}{|c|c|c|}
		\hline
		\multirow{2}{*}{\bf Parameters} & \multicolumn{2}{c|}{\bf Multi-bump model} \\
		\cline{2-3}
		& Best-fit & $95\%$ limits \\
		\hline
		$100\Omega_b$ & $2.235$ & $2.236^{+0.028}_{-0.028}   $\\
		$\Omega_{\text{cdm}}$ & $0.121$ &
		 $0.1202^{+0.0027}_{-0.0026}$\\
		$h$ & $0.6704$ & $0.673^{+0.011}_{-0.012}$\\
		$\tau_\text{reio}$ & $0.05751$ &
		$0.055^{+0.016}_{-0.015}$ \\
		$\ln 10^{10} A_s$ & $3.015$ &
		$3.042^{+0.032}_{-0.034}$ \\
		$n_s$ & $0.9644$ &
		 $0.9650^{+0.0083}_{-0.0086}$\\
		%
		%$A_{\rm I}$ & 
		$\log~A_{\rm I}$ &
		%$5.317 \times 10^{-11}$ &
		$-10.27$ &
		 $-11.9^{+1.3}_{-1.1}$\\
		%
		%$k_1/\text{Mpc}^{-1}$ & 
		$\log~k_1$ &
		%$1.806 \times 10^{-3}$ &
		$-2.74$ &
		$-2.8^{+1.6}_{-1.6}$\\
		$\Delta$ & $0.799$ &
		$0.57^{+0.42}_{-0.47} $\\
		\cline{2-3}
		$\Delta \chi^2_{\rm eff}$ & \multicolumn{2}{c|}{$-3.07$} \\
		$\ln \mathcal{B}$ & \multicolumn{2}{c|}{$0.8$}\\
		\hline		
	\end{tabular}
	\caption{\label{table:bestfit_multibump} Best-fit and $95\%$ limits of the parameters of the multi-bump model with prior $0~\leq~\Delta~\leq ~1$.}
\end{table}

\begin{figure}[!tbp]
	\centering
	\includegraphics[width=0.5\textwidth]{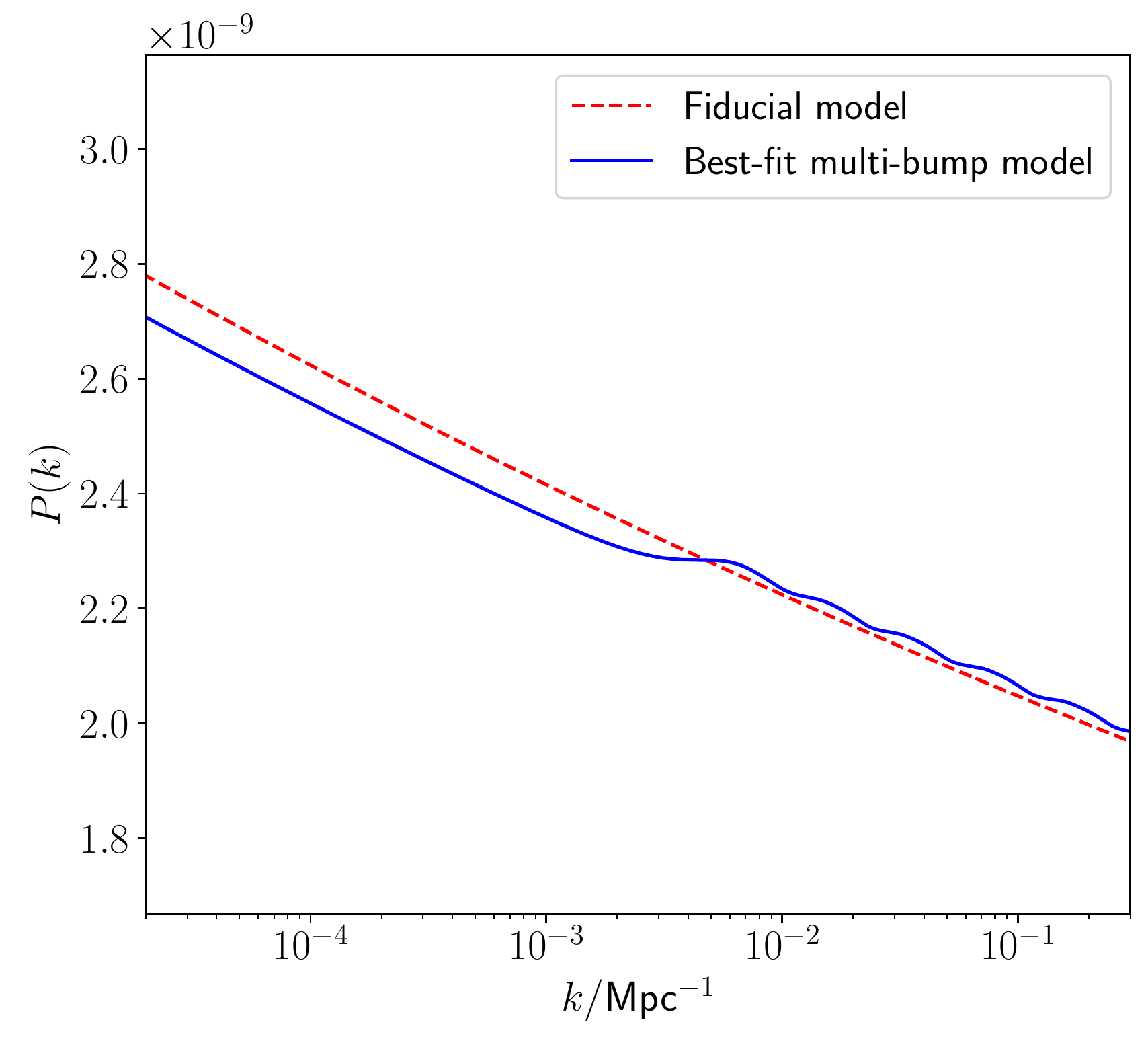}
	\caption{\label{fig:PPS_multibump}The primordial power spectrum 
	corresponding to best-fit values of the multi-bump model.
	}
\end{figure}

\begin{figure}[!t]
	\centering
	\includegraphics[width=0.6\textwidth]{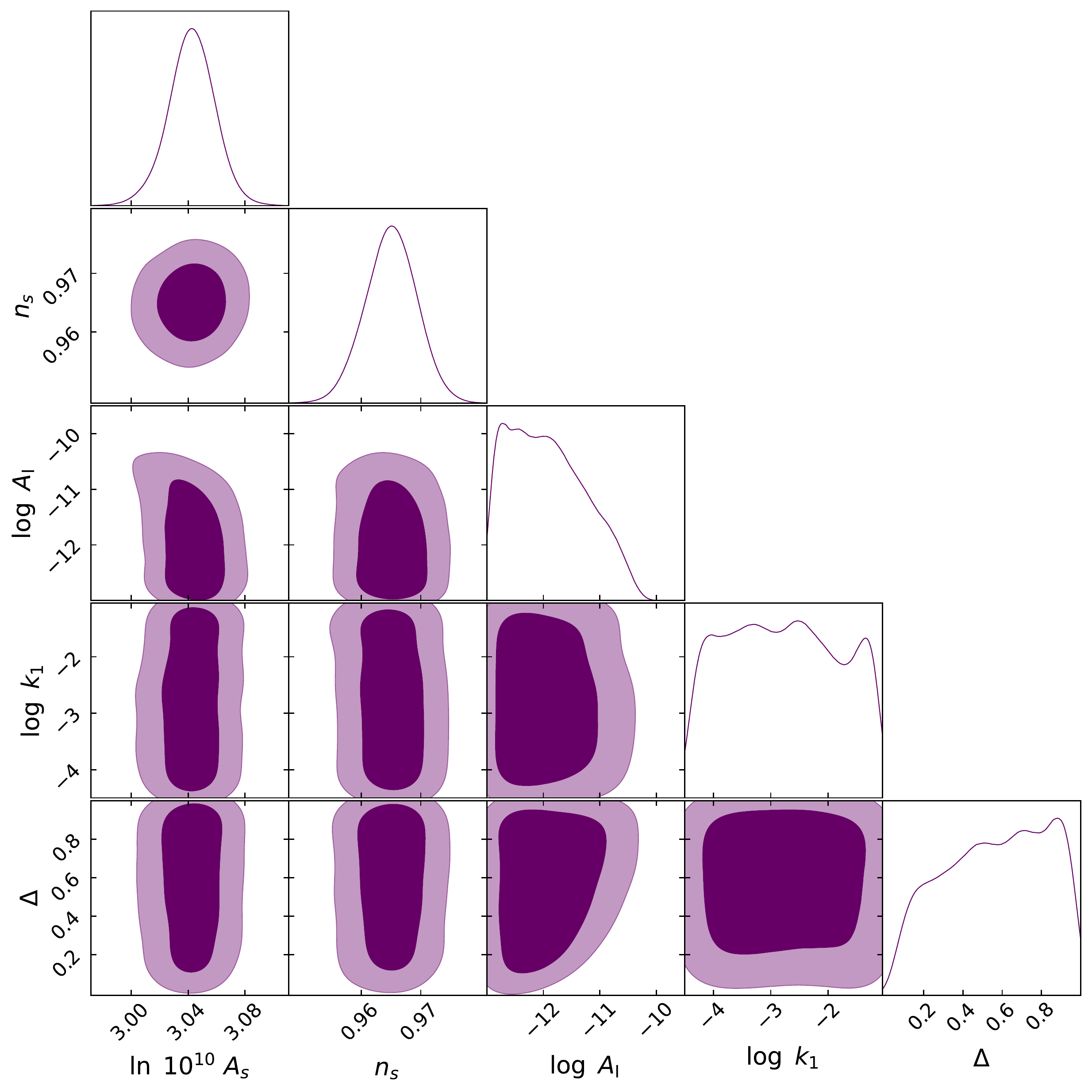}
	\caption{\label{fig:posterior_multi}The 1D and 2D marginalized 
	posterior distributions for the parameters of the multi-bump model.
	}
\end{figure}

The best-fit values and {$95\%$ limits} obtained from the Bayesian analysis of 
the multi-bump model are given in table~\ref{table:bestfit_multibump}.
%\magenta
The primordial power spectrum in eq.~\eqref{eq:PS_bump}
corresponding to best-fit values
of the parameters is shown by the solid blue curve in figure~\ref{fig:PPS_multibump}. 
For comparison, 
the power-law form for the fiducial model
is also shown (dashed red curve).  
The best-fit multi-bump model
corresponds to the particle production mechanism
commencing at\footnote{Note that, as explained earlier, the bumps start at $k_p\simeq 3.35~k_1 $%\simeq 6\times 10^{-3} \text{Mpc}^{-1}$ 
on the primordial power spectrum for the best-fit multi-bump model.} 
$k_1 \simeq 1.8 \times 10^{-3}$ Mpc$^{-1}$
and continuing at the higher $k$ - scales. 
%To visualize the multi-bumps, the 
%magnified part of the power spectrum region where particle productions
%take place is plotted in the right panel of figure~\ref{fig:PPS_multibump}.}  
%
The 1D and 2D marginalized posterior distributions
for the power spectrum parameters
are shown in figure~\ref{fig:posterior_multi}.

The last two rows of table~\ref{table:bestfit_multibump}
show that
the difference in maximum log-likelihood values
between the fiducial and the multi-bump models
is ${\Delta \chi^2_{\rm eff} \sim - 3}$. 
{Table~\ref{tab:chi2_separate} shows that the maximum 
contribution to the improvement in fitting comes from
temperature data.} 
The Bayes factor obtained for the case of the multi-bump model is 
${\ln \mathcal{B} = 0.8}$.
Compared to the single bump models we have investigated 
so far, the multi-bump model gives %\magenta
{a better
%\sout{the largest} 
improvement in the fit and the Bayes factor value.}
%As $\ln \mathcal{B}_{ij} < 1$, Jeffreys' scale 
%table~\ref{table:jeffrey} indicates that
%the data cannot distinguish the imprints
%of bump-like primordial features from the standard model.

\begin{figure}[!tp]
	\centering
	\includegraphics[width=1\textwidth]{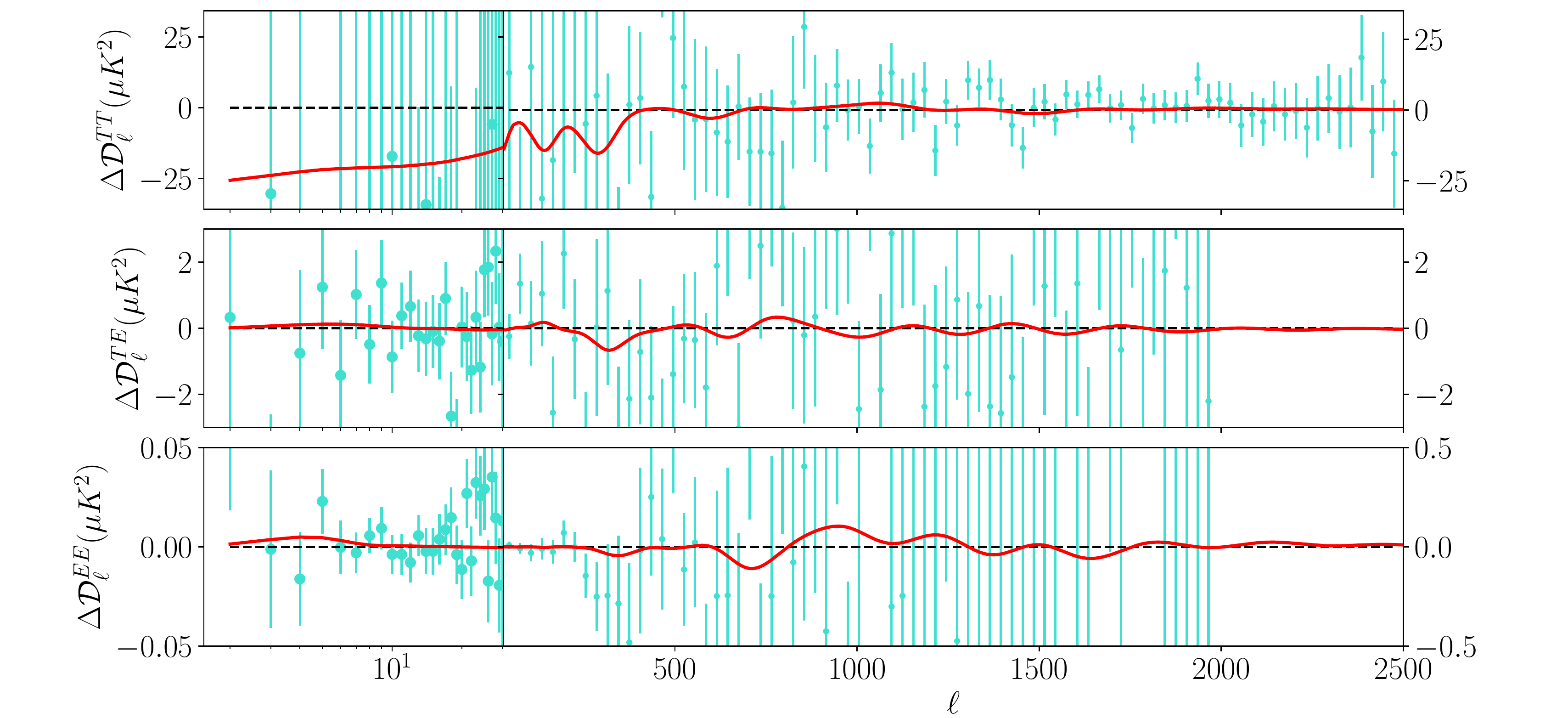}
	\caption{\label{fig:residual_multi}The residuals in CMB TT, TE, and EE power spectra of the best-fit multi-bump model with respect to the fiducial model.
	The data points with error bars are obtained 
		by subtracting the fiducial model from the 
		observed CMB power spectrum (\textit{Planck} 2018).}
\end{figure}

We plot the residuals in CMB TT, TE and EE power spectra
of the best-fit multi-bump model 
with respect to the fiducial model
in figure~\ref{fig:residual_multi}.
The data points with error bars are obtained after subtracting the best-fit fiducial model from the \textit{Planck} 2018 data. 
%\magenta
%{The improvement in the statistical
%quantities for the multi-bump case seems to be the
%improvement in the fit in the lower multipole region 
%of the CMB power spectrum.}
The improvement in the statistical quantities for the multi-bump case seems to be the improvement in the fit in the lower multipole region 
of the CMB TT power spectrum, explaining the suppression in large-scale powers followed by oscillations at higher multipoles,
{which is also evident from the $\Delta \chi^2_{\rm eff}$ values in the last row of table~\ref{tab:chi2_separate}}.
%\magenta

{Next, to determine the constraint on the theoretical model parameter $g$,
we note that the 
$95\%$ CL %\green{use either confidence level of CL}
upper bound on the amplitude of the multi-bump feature was 
$2.1 \times 10^{-11}$, or $\sim 1\%$ of $A_s$.} 
%This limit also constrains the amplitude of subdominant contributions
%to the power spectrum, i.e., $A_{\rm II} < 7 \times 10^{-13}$.
Therefore, using eq.~\eqref{eq:A1}, we find that
\begin{equation}\label{eq:gup_multi}
g < 5 \times 10^{-2}.
\end{equation}
Eq.~\eqref{eq:gup_multi} is the tightest upper bound\footnote{
%For strong couplings with $g^2>10^{-7}$, the dissipation effect may damp out  particle production \cite{Lee:2011fj}.
For coupling value $g^2>10^{-7}$, the bump in the power spectrum
may be suppressed due to the dissipation effect \cite{Lee:2011fj}.
If this is the case, the constraint on the parameter $g$
may be modified.}
obtained in our analysis and still within the natural values
(see footnote~\ref{fn:g}).
%\magenta{Describe how this compares with the constraints from the single bump case.}
%
%

%\magenta
{For completeness of the analysis,} we have also carried out the Bayesian analysis of
multi-bump models with the following prior ranges:
(i) $0\le\Delta\le3$: we obtained $\Delta \chi^2_{\rm eff}$ 
value of $-0.3$ and Bayes factor $0.6$ for this prior range,
(ii) $0 \le \Delta \le 0.5$: we obtained $\Delta \chi^2_{\rm eff}$ 
value of $-1.0$ and Bayes factor $0.7$ for this case. 
The upper bound on model parameter $g$ was consistent with
eq.~\eqref{eq:gup_multi} in both the cases.
%\magenta
{We do not discuss these cases further since 
they do not provide significantly better fits of the models to the data compared with the fiducial model.}

%
%%%%%%%%%%%%%%%%%%%%%%%%%%%%%%%%%%%%%%%%%%%%%%%%%%%%%%%%%%%%%%%%%%%%%%%
\section{Summary and Discussions}
%KF20211216 replaced {Discussions and Conclusions}
\label{sec:summary}
%%%%%%%%%%%%%%%%%%%%%%%%%%%%%%%%%%%%%%%%%%%%%%%%%%%%%%%%%%%%%%%%%%%%%%%
In this work, we investigated the
imprints of  
%for the first time?
a class of  
models 
involving particle productions during inflation 
%\cite{Barnaby:2009dd,Barnaby:2009mc,Chung:1999ve,Pearce:2017bdc}
\cite{Furuuchi:2015foh,Furuuchi:2020klq,Furuuchi:2020ery}
using the latest CMB data from \textit{Planck}. 
This class of models predicts bump-like features
in the primordial power spectrum whenever bursts of particle
production occur during inflation.
We parametrize the primordial power spectrum
using the analytical expression for the bump-like features,
including dominant and subdominant contributions
%We use the analytical form of the power spectrum
calculated in \cite{Pearce:2017bdc}.
%including the subdominant contributions.
We have carried out Bayesian analyses
to constrain one of the parameters of the theoretical model.
%To our knowledge, our work is the first
%observational analysis of signatures of 
%particle production during inflation 
%with improved analytical expressions given in \cite{Pearce:2017bdc}.
%\cite{Barnaby:2009mc,Pearce:2017bdc}. 
%With the latest CMB data from Planck 2018, 
%we have carried out Bayesian analysis
%to constrain the parameters of theoretical model. 
%Below we summarize the main results for the models with single and multiple bursts of particle production during inflation.

We summarize our results below for single burst of particle production in the following three regimes of angular scales of CMB data.
%\begin{itemize}
%    \item Single burst of particle production during inflation:
 %   The modified initial power spectrum that includes a single bump,
 %   has two additional parameters %\magenta{in comparison to $\Lambda$CDM}, namely,  
 %   the amplitude $A_{\rm I}$ and the position of the bump in co-moving wave-number $k_b~ (\text{Mpc}^{-1})$.
    %We carried out a detailed search for features throughout the 
    %scales observed by \textit{Planck}. 
        \begin{itemize}
            \item {\em Features in the large-scale (low-$\ell$) CMB data}:
            The Bayesian analysis for the search of features 
            on larger scales, i.e., prior-B 
            %($3 \times 10^{-5} < k/\text{Mpc}^{-1} < 6\times 10^{-3}$), 
            ($-4.5 < \log~k_b < -2.2$),
            returned a
            minor change in the $\chi^2$ value.
            The limit on the model parameter responsible for particle production was $g<0.1$.
            The 1D marginalized posterior distribution shows that
            multipole $\ell \lesssim 500$ in CMB data
            may accommodate primordial features of this kind
            with amplitude as large as $\sim 35\%A_s$ ($95\%$ CL).
            \item {\em Features in the small-scale (high-$\ell$) CMB data}:
            We found a marginal improvement in the 
            $\chi^2$ value ($\Delta \chi^2_{\rm eff} \sim 1$)
            when the position of the bump was confined
            in the small-scale region, i.e., prior-C %($6\times 10^{-3} < k/\text{Mpc}^{-1} < 0.1$).
            ($-2.2 < \log~k_b < -1.0$).
            %or when the bump appears in the high-$\ell$ region
            %of the CMB power spectrum.
            The constraint on the model parameter is relatively tighter 
            in the small-scale region: $g<0.06$.
            The 1D marginalized posterior distribution shows that the higher multipole CMB data may accommodate primordial 
            features with a smaller amplitude, i.e., $1.5\%A_s$ ($95\%$ CL).
            \item {\em Features in the intermediate scale CMB data}:
            We observed again a slight improvement in the fitting with prior $k_b$ having intermediate scales, i.e., prior-D 
            %($6.3\times10^{-4} < k/\text{Mpc}^{-1} < 1.0\times10^{-2} $).
            ($-3.2 < \log~k_b < -2.0$).
            The $95\%$ CL upper limit on $A_{\rm I}$ was found to be $4\%A_s$. In this case, the upper bound on the model parameter 
            was $g < 0.08$.

        \end{itemize}
        We note that the Bayes factors $\ln \mathcal{B}$ for all the single bump models are positive yet within the inconclusive range as per Jeffreys' scale (table~\ref{table:jeffrey}).
        Though we could not constrain the parameter $k_b$, our analysis with multiple prior ranges points towards 
        nearly the same best-fit location of the single bump (see figure ~\ref{fig:PPS_singleBump}).
        %\cite{Jeffreys:1939xee,Trotta:2005ar}.
    %
    %\item 
    
    The results for multiple bursts of particle production during inflation are summarized as follows:
    We found a better fit to the data in the case of multi-bump
    model when the parameter $\Delta$ that specifies the distance between the subsequent bumps was confined between $0$ and $1$, which corresponds to largely overlapping multiple bumps.
    The 1D marginalized posterior distribution returned 
    a $95\%$ CL upper limit on $A_{\rm I}$, i.e., $\sim 1\% A_s$.
    In this case, we obtained the tightest upper bound on the model parameter, $g < 0.05$.  
    The improvement in $\Delta \chi^2_{\rm eff}$ was about $-3$
    due to the better fit at low multipoles followed by bumps
    at higher multipoles. 
    The Bayes factor was 0.8, the highest one obtained in our analysis.
    Though the result does not indicate strong evidence 
    for the presence of features due to particle production
    during inflation, 
    it is interesting to note that the imprints of 
    a multi-bump model are still compatible with 
    the \textit{Planck} 2018 data. 
    %the Planck 2018 data can still accommodate
    %observational signatures of multiple bursts of particle production during inflation.
%\end{itemize}
%
Using Bayesian evidence and
%The values of Bayes factors $\ln \mathcal{B}_{ij}$
%were less than one for all the prior ranges we have considered.
%From the 
interpretations given by Jeffreys' scale
(table~\ref{table:jeffrey}),
%\cite{Jeffreys:1939xee,Trotta:2005ar},
we conclude that the presence or absence of 
bump-like primordial features is
inconclusive from the available \textit{Planck} 2018 data. 
However, we found $95\%$ CL upper limits
on the amplitudes of the primordial features $A_{\rm I}$
at different scales,
which is important for obtaining the upper bound 
on one 
of the theoretical model parameters $g$
responsible for the particle production.
%and amplitude of the subdominant contributions $A_{\rm II}$. 
%A tighter constraints on the inflation models are 
%expected from future CMB instruments with 
%improved sensitivity in temperature and 
%polarization. 
%To explore the imprints of primordial features 
%on smaller scale beyond Planck limit,
%other cosmological probes such as 
%redshifted 21~cm line would be useful. 
%Observations from 21~cm line are also expected to probe 
%the scales dominated by cosmic variance in CMB
%by tracing the matter distributions at different redshift slice.

The range of co-moving wave-numbers over which 
the temperature power spectrum is 
sensitive for primordial features, 
$0.0002 \lesssim k/\text{Mpc}^{-1} \lesssim 0.15 $,
has been covered at the cosmic variance limit 
by \textit{Planck} \cite{Akrami:2018vks}. 
The multipole range $\ell>2000$ in the temperature power spectrum, where 
experimental noise dominates the cosmic variance,
is less sensitive to the study of features as
the CMB temperature signal becomes subdominant in this region \cite{Chluba:2015bqa}.
However, the E-component polarization acts as a complementary 
dataset for searching for features in the primordial power spectrum. 
Though \textit{Planck}'s polarization data provides valuable information
at intermediate scales, it is not limited by the cosmic variance 
for small angular scales. 
Future experiments aimed at measuring CMB polarization
with improved sensitivity and broad coverage of angular scales
%\magenta {\sout{may} 
{are expected to} provide tighter constraints on the inflation models. 
%

%moved from introduction 
Primordial features arising from particle production during inflation 
are expected to leave imprints on primordial non-Gaussianity
(see, e.g.~\cite{Barnaby:2010ke}).
We do not include a discussion of non-Gaussianity here and postpone the investigation to future work.
We also do not discuss tensor perturbations in this paper since the effect of particle productions on the tensor power spectrum is expected to be small~\cite{Barnaby:2012xt,Carney:2012pk,Cook:2011hg,Senatore:2011sp,Pearce:2017bdc}. 
Any feature in the primordial power spectrum should also be encoded in other cosmological observables besides the CMB.
Following this rationale, there have been substantial efforts
to search for the signatures of primordial features in
futuristic large-scale structure surveys
\cite{LHuillier:2017lgm,Chantavat:2010vt,Chen:2016vvw,Ballardini:2016hpi,Palma:2017wxu,Ballardini:2017qwq} and
stochastic gravitational background waves 
\cite{Braglia:2020taf,Fumagalli:2020nvq}.
%\magenta
Moreover, {upcoming} %\sout{Futuristic}
redshifted 21~cm surveys can help
explore the features %\magenta
{in the primordial power spectrum} on scales inaccessible 
to other probes of matter distributions.
Recent studies have shown that future 21~cm surveys 
can improve the constraints on inflationary features
by a few orders of magnitude
\cite{Chen:2016zuu,Xu:2016kwz}.
We %\green{plan to, to make it little bit weak} 
%\sout{will} 
plan to address possible constraints on 
particle production during inflation from future 
redshifted 21~cm observations in the forthcoming work. 

\acknowledgments
We thank the anonymous referee for the 
helpful comments/suggestions.
SSN thanks Debbijoy Bhattacharya
for valuable discussions and inputs throughout
this research work.
SSN also thanks
the contributors on
Github pages \cite{baurden,brinckmann}
for helpful discussions,
in particular, 
Thejs Brinckmann,
for clarifying many questions regarding
the \texttt{MontePython} analysis.

This work was supported in part by 
Dr.~T.M.A.~Pai Ph.~D. scholarship program of 
Manipal Academy of Higher Education
and 
the Science and Engineering Research Board,
Department of Science and Technology, Government of India
under the project file number EMR/2015/002471.
Manipal Centre for
Natural Sciences, Centre of Excellence,
Manipal Academy of Higher Education 
is acknowledged for its facilities and support. \textbf{}

%%%%%%%%%%%%%%%%%%%%%%%%%%%%%%%%%%%%%%%%%%%%%%%%%%%%%
\appendix
\section{Potential degeneracy between the location of bump-like features and optical depth to reionization}\label{app:degeneracy}
%

%\section{CMB power spectra for variation in cosmological parameters and comparison with effect of bump-like features}\label{app:diff_tau}

\begin{figure}[tbp]
    \centering
    \includegraphics[width=.8\textwidth]{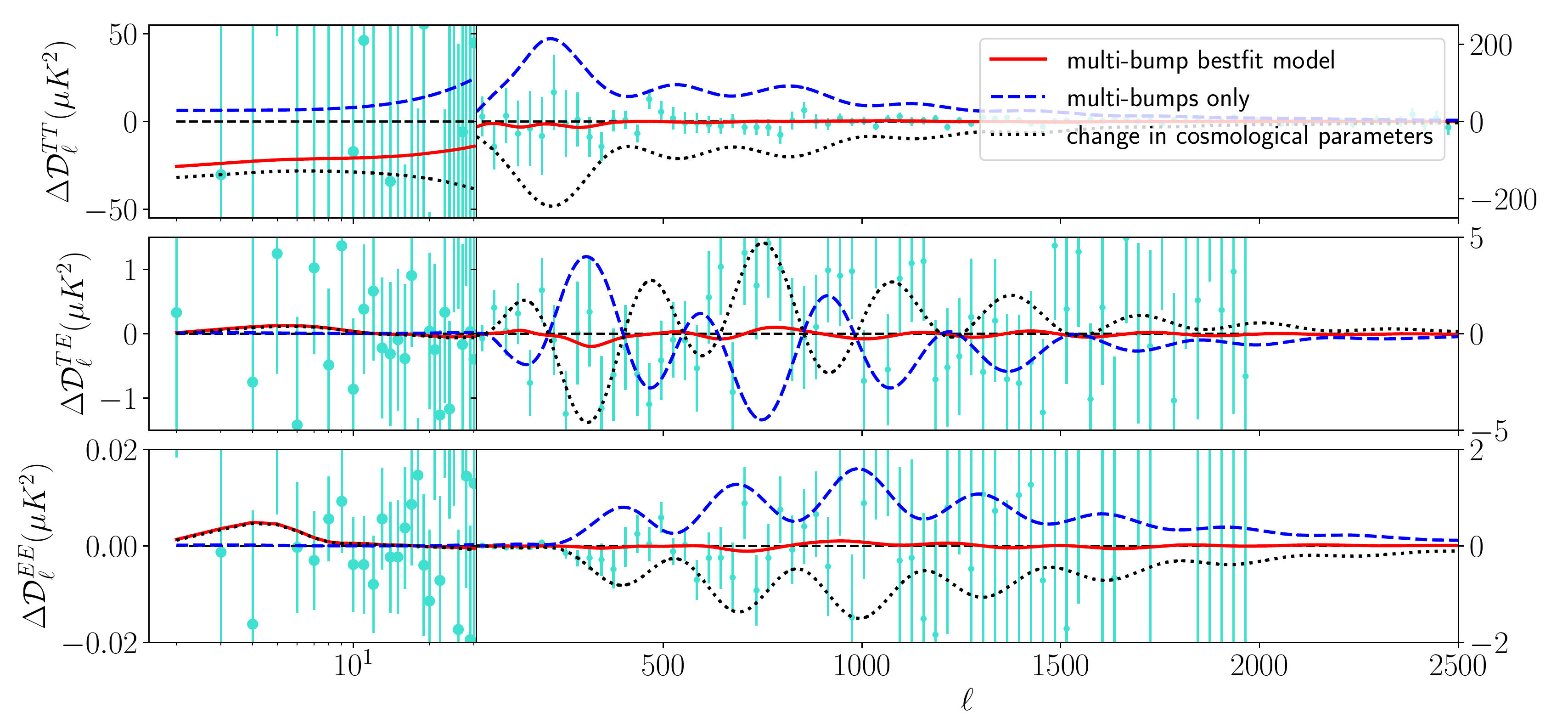}
    \caption{The residual CMB power spectra for (i) best-fit multi-bump model {\em (solid red)}, (ii) multi-bumps with cosmological parameters fixed to that of the fiducial model {\em(dashed blue)}, and
    (iii) changes in cosmological parameters without including the multi-bumps {\em(dotted black)}.}
    \label{fig:test_fit}
\end{figure}

\begin{figure}[tbp]
    \centering
    \includegraphics[width=.8\textwidth]{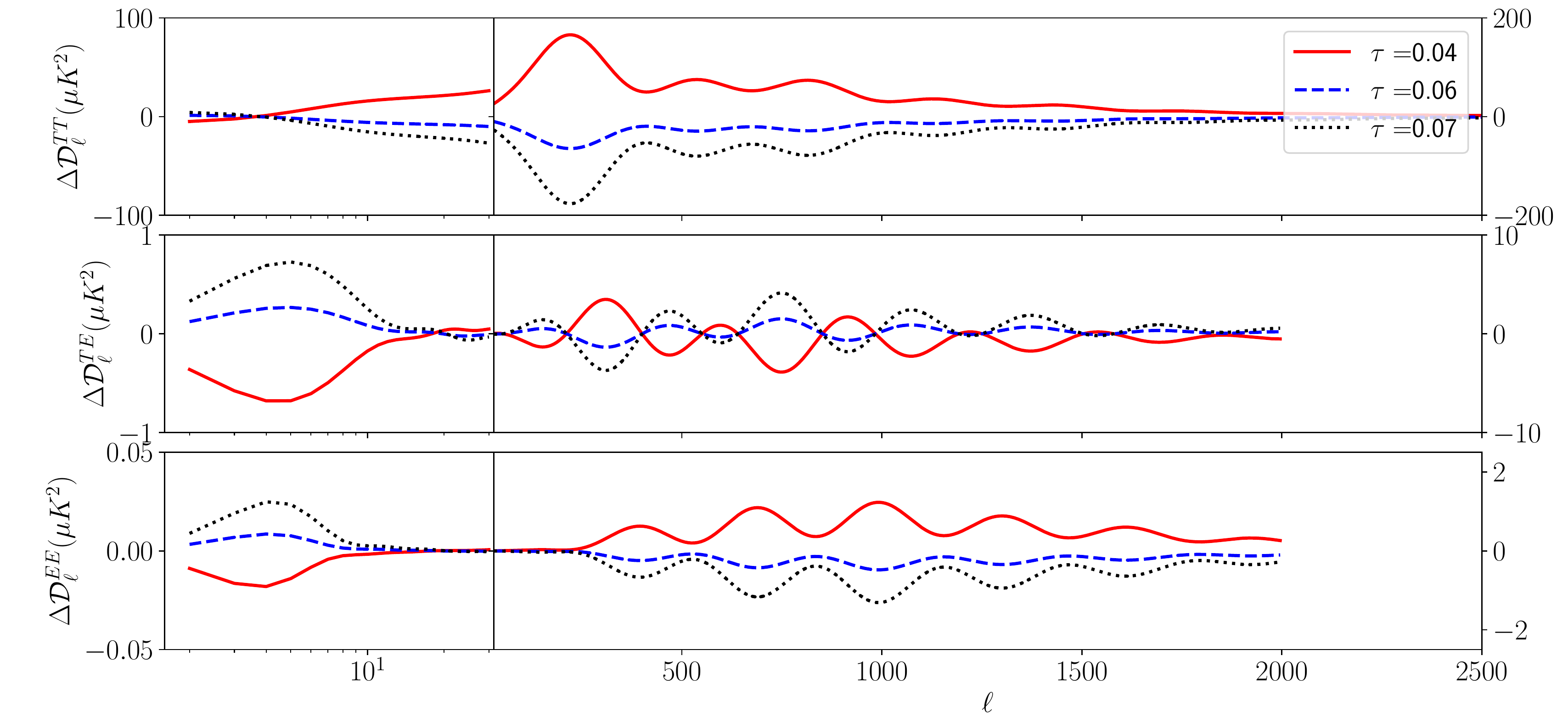}
    \caption{The residual CMB power spectra obtained by subtracting the
    \textit{Planck} best-fit model from models with different values of optical depth to reionization, $\tau$. All other cosmological parameters are fixed.}
    \label{fig:residual_diff_tau}
\end{figure}

{
In this section, we investigate the improvement in the fitting for the best-fit multi-bump model
(table~\ref{table:bestfit_multibump}) due to changes in cosmological parameters
and the addition of multi-bumps separately.
In figure~\ref{fig:test_fit}, we plot the CMB residual power spectra corresponding to
(i) best-fit multi-bump model (solid red),
(ii) contribution coming only due to the addition of multi-bumps 
- cosmological parameters are fixed to that of the fiducial model (dashed blue),
and
(iii) contribution coming only due to a shift in the cosmological parameters, from the fiducial model without including the multi-bumps (dotted black). 
We note the following:
\begin{enumerate}
    \item Lower multipole region: 
        \begin{itemize}
            \item TT power spectrum - The suppression of powers for the best-fit multi-bump model is majorly due to the shift in the cosmological parameters compared with the best-fit fiducial model. 
            \item TE and EE power spectra - The contribution to the power spectrum due to the multi-bumps in this region is negligible, and a change in cosmological parameters contributes to the excess of powers.
        \end{itemize}
    \item Higher multipole region:
        The multi-bumps start at higher $k$ on the best-fit multi-bump power spectrum, which is 
        reflected as the excess of powers in 
        TT and EE power spectra. 
        The change in cosmological parameters causes suppression of powers. 
        Therefore, the observed improvement in the fitting in this region is the combined effect of both. 
\end{enumerate}

Next, as an example, we present the effect of variations in one of the cosmological
parameters, optical depth to reionization, $\tau$.
We plot the CMB residual power spectra of TT, TE and EE correlations  for different values of $\tau$ in figure~\ref{fig:residual_diff_tau}. 
The power spectra are produced by fixing the cosmological parameters
to the \textit{Planck} best-fit value except for $\tau$ and subtracting
from them, the \textit{Planck} best-fit model with $\tau_{\rm Planck} = 0.0543$.
Varying the optical depth introduces a shift of the peak locations, in addition to varying the height of the {\em reionization bump} for the polarization power spectrum at $\ell \le 10$. 
The excess (lack) of powers in TE and EE panels at low $\ell$
is produced when $\tau > \tau_{\rm Planck}$ ($< \tau_{\rm Planck}$).} 
The variation in other cosmological parameters can shift acoustic peaks, stretch the spectrum or modulate the heights of the peaks \cite{Planck:2016tof}.
%%%%%%%%%%%%%%%%%%%%%%%
%%%%%%%%%%%%%%%%%%%%%%%%%%%%

%%%%%%%%%%%%%%%%%%%%%%%%%%
%References
%%%%%%%%%%%%%%%%%%%%%%%%%%%

\def\apj{ApJ}%
\def\mnras{MNRAS}%
\def\aap{A\&A}%
\def\apjl{ApJ}
\def\aj{AJ}
\def\physrep{PhR}
\def\apjs{ApJS}
\def\jcap{JCAP}
\def\pasa{PASA}
\def\pasj{PASJ}
\def\nat{Natur}
\def\apss{Ap\&SS}
\def\araa{ARA\&A}
\def\aaps{A\&AS}
\def\ssr{Space Sci. Rev.}
\def\pasp{PASP}
\def\na{New A}

\bibliography{CMB_constraints_ref.bib}
\bibliographystyle{JHEP}

\end{document}